\documentclass[fleqn,usenatbib]{mnras}

\usepackage[utf8]{inputenc}
\usepackage{amsmath}
\usepackage{graphicx}
\usepackage{epstopdf}
\usepackage{graphicx, color}
\usepackage{textcomp}
\usepackage{bm}
\usepackage{soul}
\usepackage{amssymb}
\usepackage{hyperref}
\usepackage{blindtext} 
\usepackage{subfig}
\usepackage{tabularx}
\usepackage{threeparttable}
\newcommand{\eqb}{\begin{eqnarray}}
\newcommand{\eqe}{\end{eqnarray}}
\newcommand{\sth}{\sigma_{\rm T}}

\newcommand{\tesc}{t_{\rm esc}}

\newcommand{\lp}{\ell_{\rm p}}
\newcommand{\lpcrss}{\ell_{\rm p, ss}}
\newcommand{\lpcr}{\ell_{\rm p,cr}}
\newcommand{\lph}{\ell_{\gamma}}

\newcommand{\gmx}{\gamma_{\max}}
\newcommand{\gmn}{\gamma_{\min}}

\newcommand{\mpr}{m_{\rm p}}

\title[Variability in hadronic supercriticality]
{Patterns of variability in supercritical hadronic systems}
\author[]{M. Petropoulou$^{1}$\thanks{E-mail:m.petropoulou@astro.princeton.edu} \& A. Mastichiadis$^{2}$ \\
$^{1}$Department of Astrophysical Sciences, Princeton University, 4 Ivy Lane, Princeton, NJ, 08540, USA \\
$^{2}$Department of Physics, National and Kapodistrian University of Athens, Panepistimiopolis, GR 15783 Zografos, Greece\\
}
\begin{document}
\date{Received.../Accepted...}

\pagerange{\pageref{firstpage}--\pageref{lastpage}} \pubyear{2018}

\maketitle

\label{firstpage}
\begin{abstract}

A unique and often overlooked property of a source loaded with relativistic protons is 
that it can become supercritical, i.e. it can undergo an abrupt transition from a radiatively inefficient to a radiatively efficient state
once its proton energy density exceeds a certain  threshold. In this paper, we investigate the temporal variability of hadronic systems in this hardly explored regime. We show that
there exists a range of proton densities that prevent the system 
from reaching a steady state, but drive it instead  in a quasi-periodic mode. The escaping radiation then exhibits  limit cycles, even if all physical parameters are held constant in time.  We extend our analysis to cases where the proton injection rate varies with time and explore the variability patterns of escaping radiation as the system moves in and out from the supercritical regime.  
We examine the relevance of  our results to the variability of the prompt gamma-ray burst emission and show that, at least on a phenomenological level, some interesting analogies exist. 
 
\end{abstract}

\section{Introduction}
\label{intro}

Hadronic models for high-energy emitting sources, such as active galactic nuclei {(AGN) \citep[e.g.][]{sikora87, mannheim93, rachen98, aharonian00, mueckeprotheroe01} gamma-ray bursts (GRB) \citep[][]{waxman97,boettcherdermer98, kazanas02, murasenagataki06, asano07}, and binary systems \citep[e.g.][]{romero03, romero05, vieyro12},
constitute an alternative to the widely accepted leptonic models. 
An often overlooked, yet unique, property of sources loaded with relativistic protons
is the abrupt transition from a low to a high radiative efficiency regime. The onset of this transition, which is coined 
``proton supercriticality'' \citep{kirkmast92},
occurs when the energy density of protons in the source exceeds a certain threshold (critical value). 
The energy stored into protons is then released fast  (i.e., within a few dynamical timescales) and with a high radiative output. 
As a consequence, the number of photons increases abruptly in the source, leading to additional cooling of the protons. 

The temporal evolution of the proton distribution in the source strongly depends on that of the photons and {\it vice versa}. Despite the complexity
of the radiative processes that are involved in this coupling (e.g., photopion and photopair production processes), it turns out that the equations
governing their temporal evolution are of the Lotka-Volterra type \citep{petromast12b}. Even if energy is provided into the hadronic system at a constant rate, this may exhibit intrinsic variability in the form of quasi-periodic oscillations whose frequency increases as the system is driven deeper into the supercriticality. In  more realitistic situations, the energy injection rate may also vary with time pushing the system in and out of the supercritical regime. This can lead to the production of structured light curves which may have some relevance to those observed from high-energy emitting sources. 

The response of the hadronic system to time-dependent variations of the energy injection was investigated in the context of blazars, a subclass of AGN with jets pointing towards the observer \citep{mpd13}. For the relevant parameters, the system always lied in the so-called subcritical regime that is characterised by low radiative efficiency. The response of the hadronic system to time-dependent variations of the energy injection in the supercritical regime remains, however, hardly explored. Aim of the present paper is to investigate the temporal properties of the hadronic system  in the supercritical regime by studying the photon light curves and their respective Fourier transforms in cases where the energy injection is also time variable. 

The present paper is structured as follows:
 In Section \ref{sec:model} we present the physical conditions of the model and outline our methods.  
In Section \ref{sec:temporal}  we first explore the transition between
subcritical and supercritical states in the case of constant proton injection and then 
we examine the case of variable injection. In Section \ref{sec:grb} we produce light curves for GRB-like  parameters and show that there are some analogies between our results
and the observed GRB light curves of the prompt phase.
We conclude in Section \ref{sec:discuss} with a discussion of our results.
 
\section{Proton supercriticalities}\label{sec:model}
In what follows we present the principles of the hadronic supercriticality and we describe briefly the numerical code that we will use.
\subsection{The concept}
Relativistic protons inside compact sources can interact with 
ambient magnetic fields and radiation through synchrotron,  photopair (Bethe-Heitler), and photopion production processes. 
Relativistic electron-positron pairs are injected in the source via the Bethe-Heitler process, whereas 
photopion interactions result in the production of many unstable charged particles, like
pions ($\pi^{\pm}, \pi^0$), muons ($\mu^{\pm}$) and kaons ($K^{\pm}, K^0$), which decay into lighter particles. For example, 
the decay of $\pi^{\pm}$ results in the injection of secondary relativistic 
electron-positron pairs: $\pi^{\pm} \rightarrow \mu^{\pm}+\nu_{\mu}(\bar{\nu}_{\mu})$, $\mu^{\pm}\rightarrow e^{\pm}+\bar{\nu}_{\mu}(\nu_{\mu})+\nu_{\rm e}$($\bar{\nu}_{\rm e}$), while $\pi^0$ decay into very high energy $\gamma$-rays. Relativistic neutrons are also a by-product of photohadronic collisions and they may also interact with ambient photons before they decay, as their lifetime is prolonged by a factor equal to their Lorentz factor. 

The injection of secondary relativistic leptons increases significantly the number of synchrotron photons in the source, while protons do not typically cool fast. The combination of these two effects may give rise to hadronic supercriticalities, while leading to some interesting temporal behaviour. This was first pointed by  \cite{sternsvensson91} with the help of a Monte Carlo code that treated proton losses and emission from its secondaries. Later \cite{mastkirk95} found a similar behaviour by numerically solving a system of three time-dependent coupled kinetic equations  describing the evolution of protons, electrons,  and photons in a spherical
volume entrained by a magnetic field. The numerical findings of \cite{mastkirk95} verified that  
relativistic protons inside a volume can become supercritical once a feedback and a marginal stability criterion are simultaneously satisfied, as first demonstrated by the stability analysis of \cite{kirkmast92}.  

The basic concept of the hadronic supercriticality is simple and 
can be understood as follows. 
The injection of relativistic protons inside a source will 
result in an increase of their energy density as long as protons do not cool efficiently or escape from the system. When the proton energy density becomes sufficiently large, protons may become targets of their own radiation (either directly from their synchrotron radiation or indirectly from the radiation of their secondaries). This leads to an exponentiation of the number of photons in the source and eventually to very fast proton losses due to the proton-photon coupling through photopair and photopion interactions.  The radiative feedback on the protons is feasible, if the produced   photons are energetic enough to meet the
threshold conditions of the two aforementioned photohadronic processes. These, in turn, translate to conditions between  
the proton maximum energy and the strength of the magnetic field \citep{kirkmast92}.
 
\cite{petromast12b} derived a simplified network of equations describing the evolution of protons and photons when both species are coupled via photohadronic interactions while energy is  being pumped into the system at a constant rate.  By performing linear stability analysis to the system of equations, they showed that the behaviour of the system  can  fall generally  into two broad categories: 
\begin{enumerate}
 \item for low enough proton densities (subcritical regime), a steady state (equilibrium) is established after a few 
 dynamical timescales; the photon and proton densities do not evolve with time.
 \item for high enough proton densities (supercritical regime), the temporal evolution of the proton and photon densities resembles that of a  ``prey-predator'' system and the light curve exhibitis quasi-periodic oscillations (limit cycles).
\end{enumerate}

\subsection{The numerical code}
The study of hadronic supercriticalities and of the system's temporal evolution requires
a numerical approach which can treat radiative feedback effects in a self-consistent way. For this purpose, 
we have adopted a numerical code that solves a set of five time-dependent, energy-conserving kinetic equations, describing the evolution of the energy
distributions of the stable particle populations in the source \citep{dmpr12}. The code is ideal for addressing various topics, such as the efficiency of energy conversion from protons to photons and neutrino \citep[see also][]{pgd14, pdmg14, petro15}, the relation between their respective spectral shapes, as well as the temporal behaviour of the system, a feature that is central for the present paper.

The input parameters to the code are the source size $R$, the magnetic field strength $B$, the particle escape timescale, $\tesc \ge R/c$, and the characteristics of particles (protons and primary electrons) at injection, i.e., the power-law slope, the minimum and maximum Lorentz factors  of the distributions, as well as the injection compactness $\ell_{p,e}$. This is a dimensionless measure of the particle luminosity $L_{\rm e,p}$  and is defined as $\ell_{\rm e,p} =\sth L_{\rm e,p}/4 \pi R m_{\rm e,p} c^3$.

To delineate the role of protons from that of the primary (i.e., accelerated) electrons to the  system's evolution, we choose $L_{\rm e} \ll L_{\rm p}$, i.e. primary electrons make a negligible contribution to the radiation energy density of the system at all times.

\section{Variability patterns}
\label{sec:temporal}  

\subsection{Constant proton injection}\label{sec:const}
The variability expected in the case of constant injection in the supercritical regime has been presented by \cite{petromast12b}. Here,  
we repeat their main findings through an indicative example, which will serve as the 
basis of this study. 

The parameters that bring the system in a supercritical state
are not known {\sl{a priori}}. \cite{kirkmast92} have provided some analytical
relations that have been later verified numerically \citep{mastetal05}, but these do not constitute a comprehensive set of the supercriticality conditions. For any set of parameter values,  one has to determine numerically the ``critical'' proton compactness ($\lpcr$) above which the system enters the supercritical regime. 

We, therefore, performed a suite of numerical runs to determine $\lpcr$, where we were increasing $\lp$ by a factor of two over its previous value, while keeping all the other free parameters of the problem fixed ($B=3\times10^3$~G, $R=10^{12}$~cm, $\gmx=3\times10^7$,
$\gmn=1$, $s=2$  and $\tesc=1000\ R/c$). 
\begin{figure}
\centering
 \includegraphics[width=0.44\textwidth]{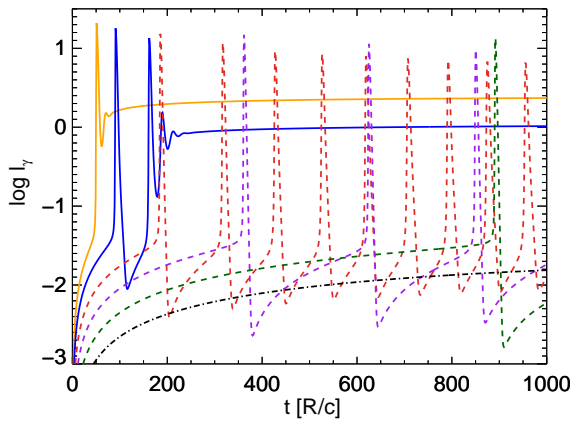}
 \caption{Plot of the photon compactness (in logarithm) versus time (measured
 in units of $R/c$) for a sequence of proton injection compactnesses that 
 differ from each other by a factor of two. Other parameters used are: $B=3\times10^3$~G, $R=10^{12}$~cm, $\gmx=3\times10^7$,
$\gmn=1$, $s=2$  and $\tesc=1000\ R/c$. For this particular set, we have also determined numerically the characteristic values of the proton compactness, namely $\lpcr\simeq 6.3\times10^{-5}$ and $\lpcrss \simeq 4\times10^{-4}$.}
\label{Qcon}
\end{figure}
Figure \ref{Qcon} presents a typical set of light curves from our simulations. For low enough proton injection values, the light curve steadily increases until it reaches a constant value (black dash-dotted line).  The system goes into
a steady state without entering the supercritical regime. The proton density does not build inside the source to high enough values as to trigger self-sustained photohadronic reactions with the proton synchrotron photons. The main energy loss mechanism is proton synchrotron radiation which is, for the given set of initial parameters, inefficient. Consequently,  the photon luminosity is low and the system is radiatively inefficient in that only a small fraction of the injected proton power is transformed into radiation.  For sufficiently high values of the proton compactness, the  above picture changes drastically and the system enters into the supercritical regime (dashed lines). The transition is manifested by flares that release a substantial part of the energy stored in relativistic protons. During the flares the photon luminosity increases, in the present example, by a factor of $\sim300$. As the number of photons increases exponentially, protons undergo severe energy losses and move back into the subcritical regime leading photons back to low state. The flares can therefore last only for  a few dynamical timescales.

The radiative efficiency increases as most of the proton energy is turned into radiation, albeit in an
intermittent way. We thus determined $\lpcr$ to be the lowest value of the proton compactness
that produces at least one photon flare within a time interval equal to $\tesc$\footnote{In the absence of proton losses, the system would reach a steady state within the same time interval.}. As $\lp$ increases beyond $\lpcr$, the system is driven deeper into the supercritical regime. This is manifested by the increased number  of photon flares and by the shorter time interval between successive outbursts. However, if $\lp$ exceeds another characteristic values, which we shall define as $\lpcrss$,  the system reaches once again
a steady state. This may happen either through a series of dumped oscillations or, as in the
present case, via an initial outburst  (solid blue line). A further increase of $\lp$  does not change the
basic picture as the system reaches still a steady state but now at earlier times (solid orange line). Note that in this supercritical steady state 
the system is efficient: the proton luminosity is radiated away in the form of photons and neutrinos
via the photohadronic processes that become increasingly important as the proton injection compactness increases. 

From the above analysis, it is evident that
for a given set of parameters one can find two values of $\lp$ 
that determine fully the hadronic system's temporal behaviour and efficiency.
On the one hand, the system is driven to the supercritical regime
and limit cycle behaviour for $\lp \ge \lpcr$, and, on the other, if $\lp$ is to exceed the value $\lpcrss$, then any intrinsic variability is suppressed and the system reaches once again a steady state. In other words, the values  of the proton injection compactness lying
in the range $(\lpcr, \lpcrss)$ act oppositely to a frequency filter, since they allow the emergence of  intrinsically periodic signals.
In summary, we identify three regimes based on the value of the proton compactness:
\begin{itemize}
\item low zone, for $\lp < \lpcr$. The system is subcritical, shows no intrinsic variability, and has low radiative efficiency.
\item intermediate zone, for $\lpcr < \lp \le \lpcrss$. The photon luminosity varies quasi-periodically, even if the source of particle injection is constant in time. The radiative efficiency is high. 
\item high zone, for  $\lp > \lpcrss$. The radiative efficiency is high, but any intrinsic quasi-periodicity disappears.
\end{itemize}
Note that the above is only an illustrative example that will help extend our analysis to time variable
proton injection. A more comprehensive study on the dependence of $\lpcr$, $\lpcrss$ 
on the parameters of the problem, such as the magnetic field and $\gmx$ will be the subject of a forthcoming publication.
\begin{figure}
\centering
 \includegraphics[width=0.44\textwidth]{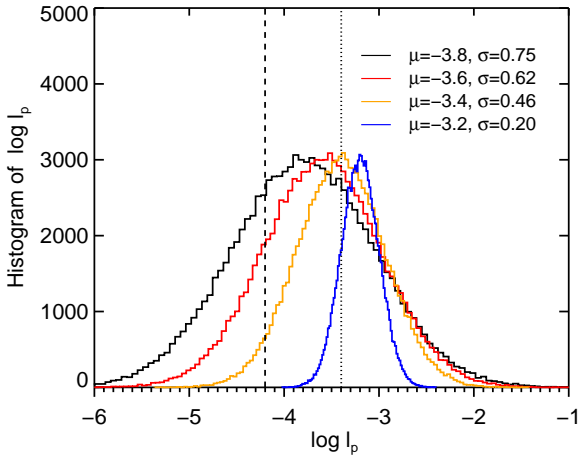}
 \caption{Histogram of the $10^5$ random numbers following normal distributions with different means ($\mu$) and dispersions ($\sigma$) -- see inset legend. $N=1000$ numbers were drawn from each distribution to simulate the proton injection  compactness as a function of time according to eq.~(\ref{eq:white}). The values of $\mu, \sigma$ were appropriately chosen to conserve the total injected
 energy. For reference, we also mark on the plot the values of $\lpcr$ and $\lpcrss$ (dashed and dotted lines, respectively). 
 }
 \label{histo}
\end{figure}
\subsection{Variable proton injection}\label{sec:var}
As there is no {\sl a priori} reason for assuming 
a constant particle injection rate in the emission region, here 
we study the effects of a variable source of injection
on the variability properties of the hadronic system.
For reasons of concreteness,  we adopt the same parameter set as the one used in the previous section, which we keep fixed from this point on.

We assume that $\lp$ shows small amplitude variations
on the light crossing timescale of the source, while its power spectral density (PSD) 
is described by a power law, i.e. $P_{\rm f} \propto f^{-\alpha}$.
In particular, we study the properties of the photon light curves 
and their PSDs for the characteristic cases of $\alpha=0$ (white noise)
and $\alpha=2$ (red noise). 

\begin{figure*}
\begin{tabular}{c c }
\subfloat[]{\includegraphics[width=0.44\textwidth]{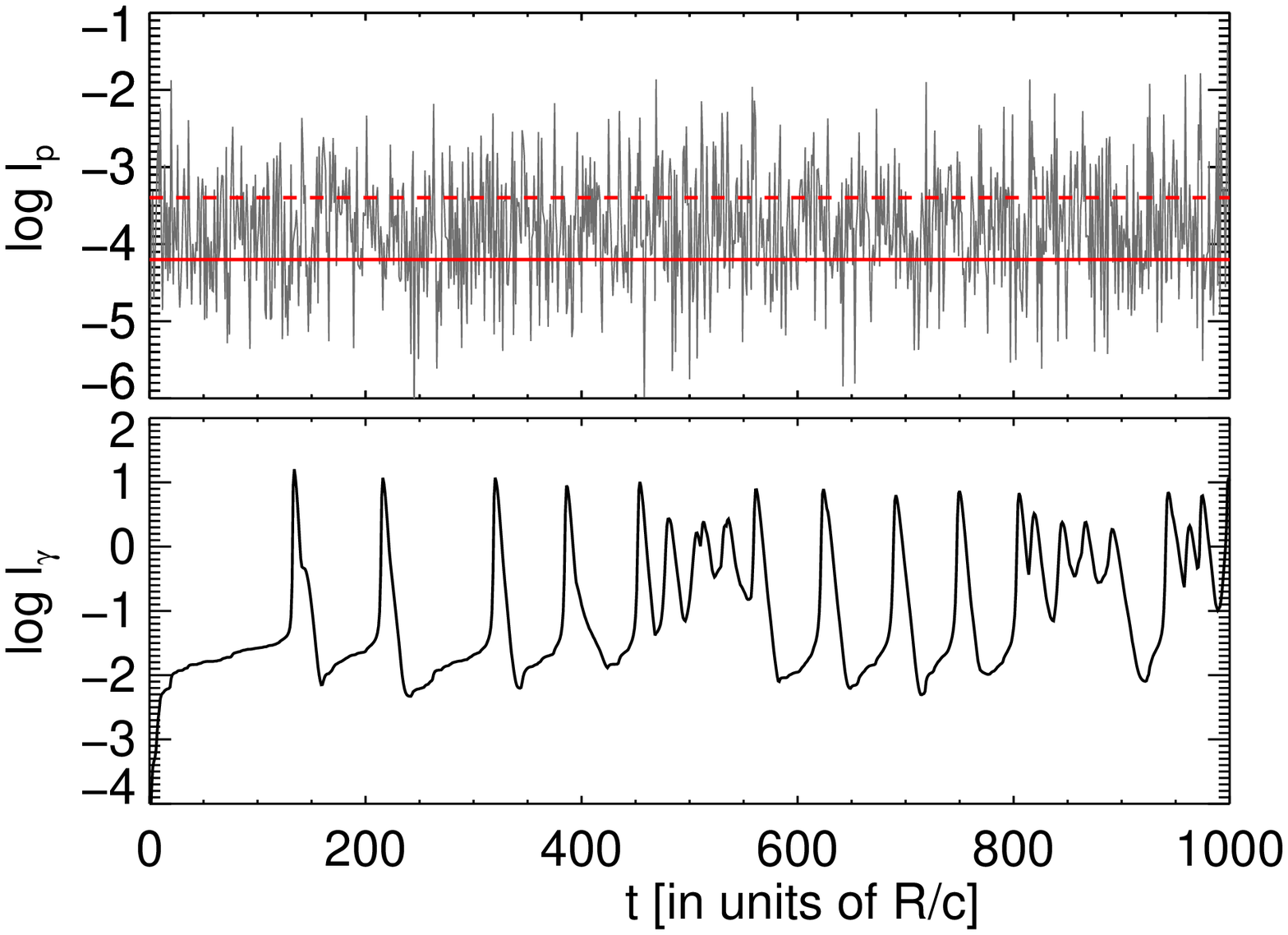}} &
\subfloat[]{\includegraphics[width=0.44\textwidth]{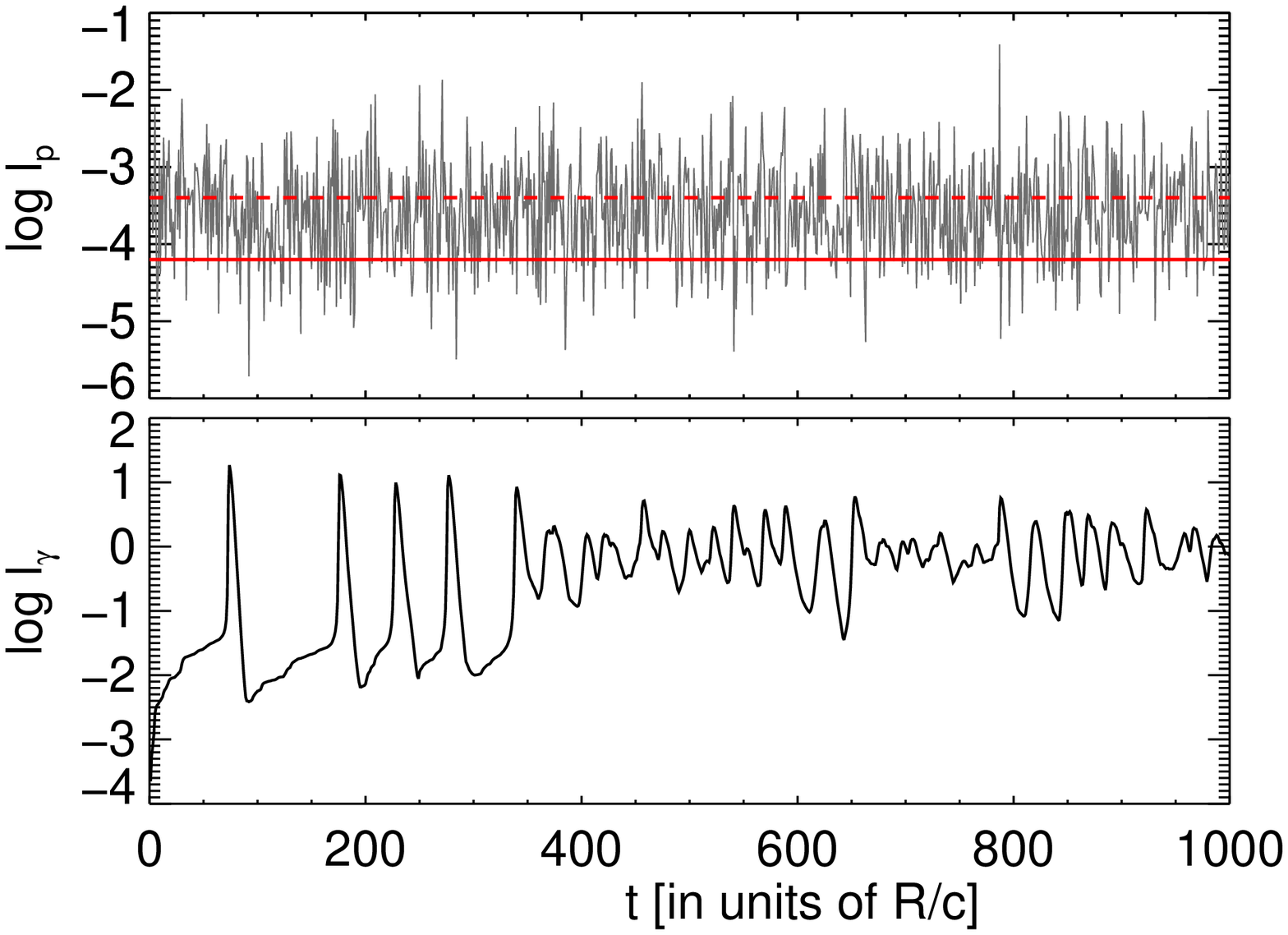}} \\
 \subfloat[]{\includegraphics[width=0.44\textwidth]{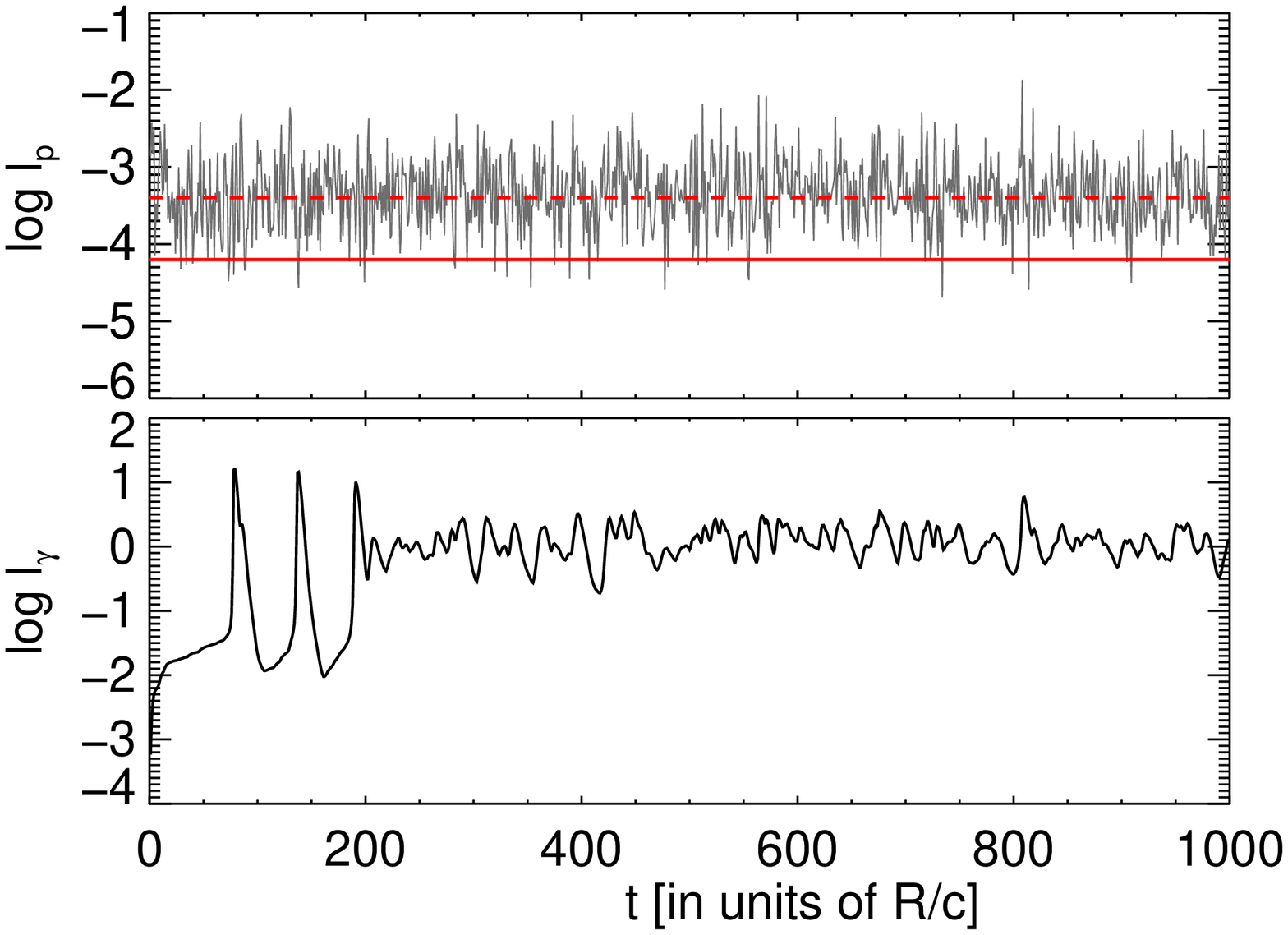}} &
\subfloat[]{\includegraphics[width=0.44\textwidth]{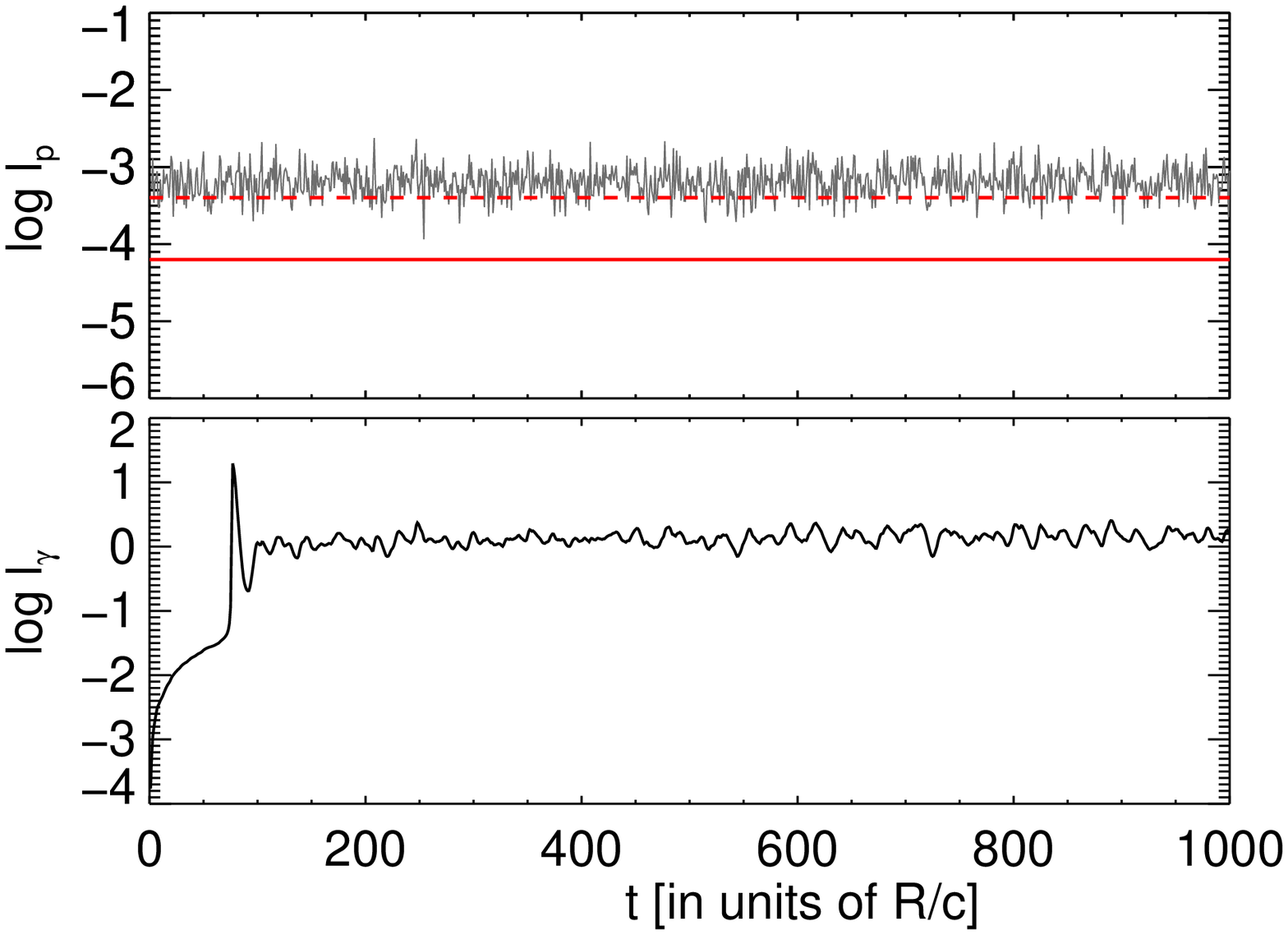}}
\end{tabular}
\caption{Plot of the proton (top) and photon (bottom) compactnesses (in logarithm) as a function of time (in $R/c$ units)
obtained for different normal distributions of random numbers. 
Results for $(\mu, \sigma)=(-3.8, 0.75), (-3.6, 0.62), (-3.4,0.46)$, and $(-3.2,0.2)$ are shown in  panels (a) to (d). The horizontal red lines indicate  the values of $\lpcr$ (solid) and $\lpcrss$ (dashed), respectively. 
}
\label{white}
\end{figure*} 
\subsubsection{White noise}\label{sec:white}
At each time step $\tau_{i+1}\equiv ct_{i+1}/R$, the logarithm of the proton compactness is determined by its value at the previous time step as:
\eqb
g_{\rm i+1} & = &  g_{\rm i} + r_{\rm i},\ i=1,\cdots, N-1 
\label{eq:white}
\eqe
where $g$ stands for $\log\lp$, $N=1000$, and $r$ is a random number drawn from a normal distribution with mean $\mu$ and dispersion $\sigma$.
We performed four numerical simulations with the same initial conditions (i.e., $\tau_1 =0$ and  $g_1=-4.4$) but with different values of $(\mu, \sigma)$  -- for details, see the caption of Fig.~\ref{histo}. 
The amount of time that $\lp$ spends in the three regimes defined in Sect.~\ref{sec:const} is different
for the four cases. However,  by choosing appropriately the characteristics of the normal distribution, i.e. its mean 
$\mu$ and its dispersion $\sigma$, we ensure that the total injected energy into the system ($e_{\rm inj}=\int {\rm d}\tau \ \lp(\tau)$) is the same in all cases. 

The photon compactness as a function of time obtained for the different time profiles for $\lp$ described above
is shown in Fig.~\ref{white}. In all cases we find that the shortest time variations of $\lp$  (i.e., on one crossing timescale) are absent from the photon light curves, even when  the system is driven in the supercritical regime where cooling of relativistic protons, at least, is efficient.
Although the amount of energy being provided to the system is approximately equal in all four cases,
the way that the energy is being released differs significantly: through large amplitude flares (panel a)
to small amplitude flares that follow on average the variations of the injection (panels c and d).
The radiative efficiency of the system also increases from panel (a) to panel (d), since the time that $\lp$ lies above $\lpcr$ becomes increasingly larger.

\subsubsection{Red noise}\label{sec:red}
As a next step, we introduce a variable injection that can be described by a
red noise process, i.e., less power is being injected at shorter
timescales. This case where all timescales are not equivalent might be more
relevant to astrophysical sources. Similarly to the previous paragraph, the logarithm of the proton injection compactness is modelled
as:
\eqb
g_{\rm i+1} & = &  g_{\rm i} + A(-1)^{[u_{\rm i}]},\ i=1,\cdots, N-1 
\label{eq:red}
\eqe
where $u$ is a random number uniformly distributed in (0,10), $[\cdots]$ represents the integer
part of a number and $A$ is the amplitude of the variation.
Given a series of random numbers, the parameter $A$ determines 
for how long $\lp$ lies in the low, intermediate, and high regimes, respectively.
We start from $g_1=-4.4$ using the same series of random numbers for
four values of $A$ ($0.05$, $0.075$, $0.1$, and  $0.125$).
In contrast to the numerical simulations presented in Sect.~\ref{sec:white}, the total injected energy
increases between successive runs of larger $A$ values. In addition, the system spends more time in the supercritical regime (including the intermediate and high zones), as $A$ increases. 

The light curves are presented in Fig.~\ref{red} for increasing values of $A$ from  panel (a) to (d). Two major outbursts are produced in the first case (panel a) that carry a substantial amount of the energy stored in protons. These outbursts appear almost periodically and the temporal behaviour of the system is similar to the limit cycle behaviour discussed in the case of constant particle injection. The photon luminosity increases by a factor of $\sim 1000$  during the two outbursts  and it can exceed, for a few crossing times, the one injected in protons. 
If  the proton's compactness does not exceed for a long time $\lpcrss$, the outbursts will have a duration of only a 
few $R/c$ as the photons will keep rising until the protons cool in such a degree as to drive the system to the  subcritical regime. 
Meanwhile, photons leave the system and the burst duration is  essentially defined
by $R/c$. In this case the outbursts are single spikes. 
If the proton synchrotron cooling timescale is long, like in this specific example,
all small scale structure originally seen in the proton injection curve is wiped out 
from the photon light curve. 

Panel (b) demonstrates a case where the system has a higher total energy budget and therefore enters
the supercritical regime more often. The bursts that appeared 
as single spikes in panel (a) now exhibit sub-structure that mimics that of the proton injection temporal profile. The appearance of substructure  happens when the system entered the high zone ($\lp>\lpcrss$) for long enough time as to
suppress any intrinsic quasi-periodicity. The same holds if the system returns to the subcritical state,
although the photon luminosity will be much lower in this case. The total light curve shown in panel (b) consists of structured flares obtained  from a prolonged exposition of the  system to supercritical regime and single spiked flares obtained for $\lpcr< \lp < \lpcrss$, same 
as in panel (a). Finally in panels (c) and (d) the above become even more pronounced, and the  system shows again both
structured and single bursts being more clustered together. 
\begin{figure*}
\begin{tabular}{cc}
\subfloat[]{\includegraphics[width=0.44\textwidth]{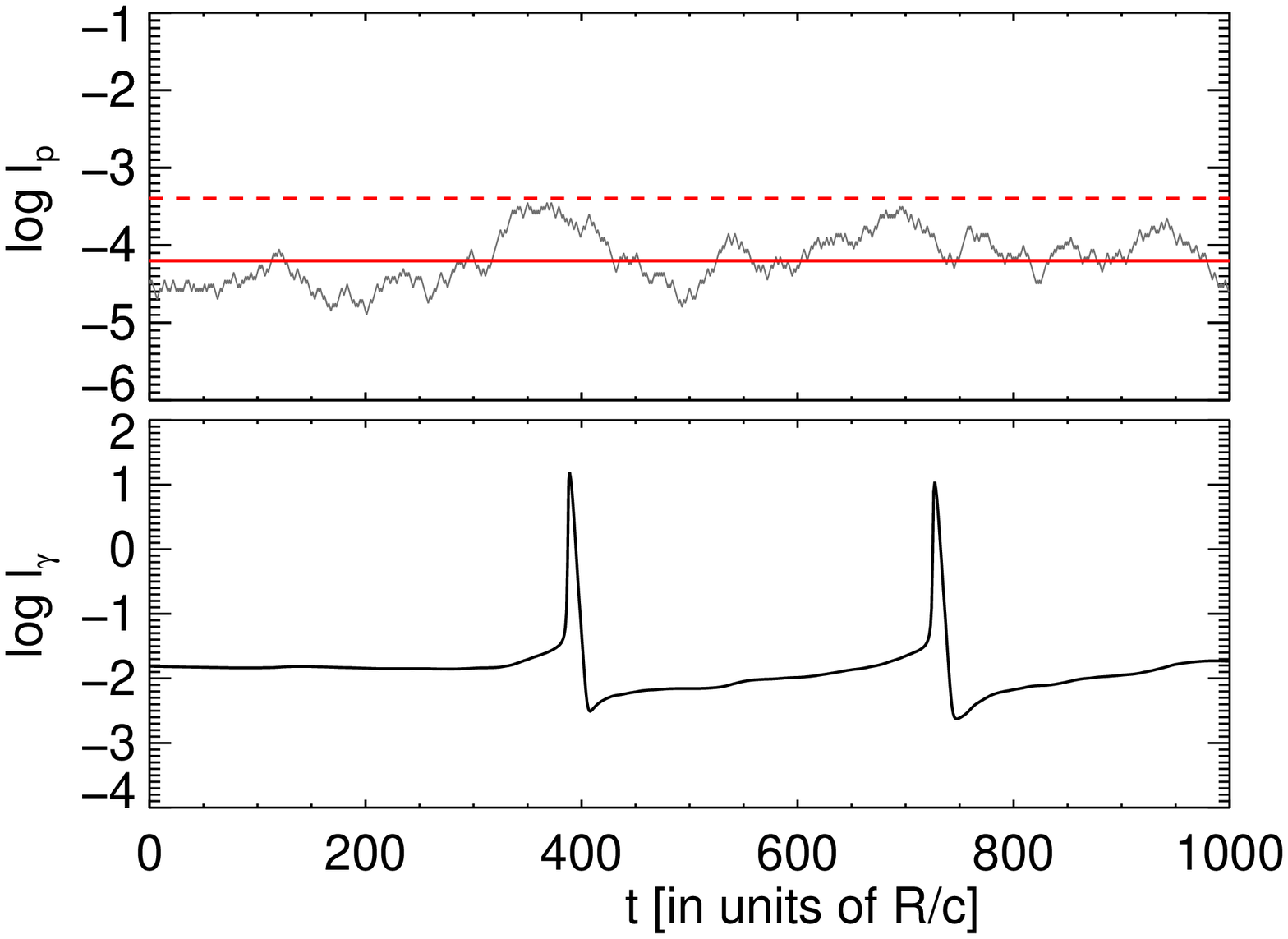}} &
\subfloat[]{\includegraphics[width=0.44\textwidth]{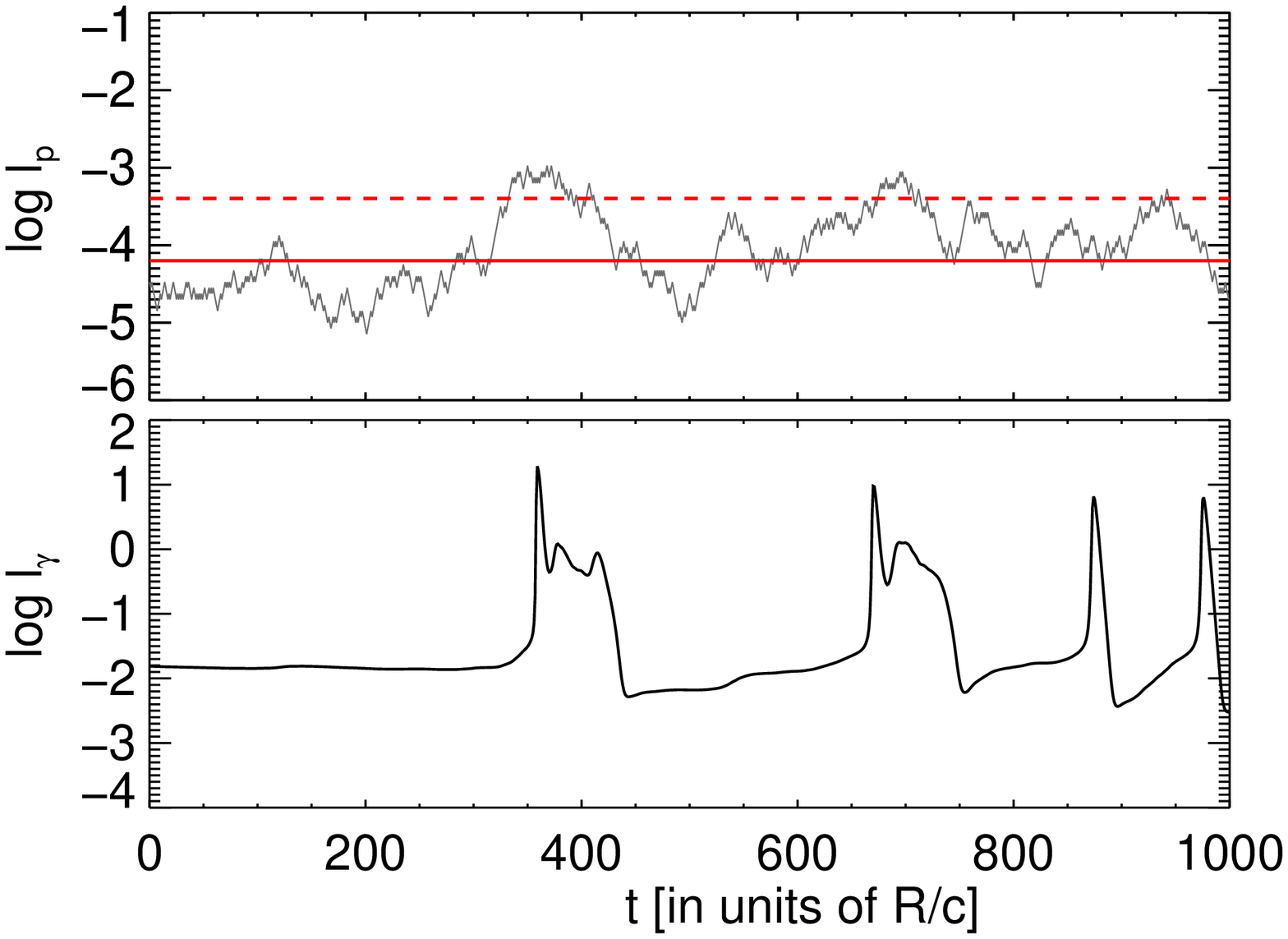}} \\
\subfloat[]{\includegraphics[width=0.44\textwidth]{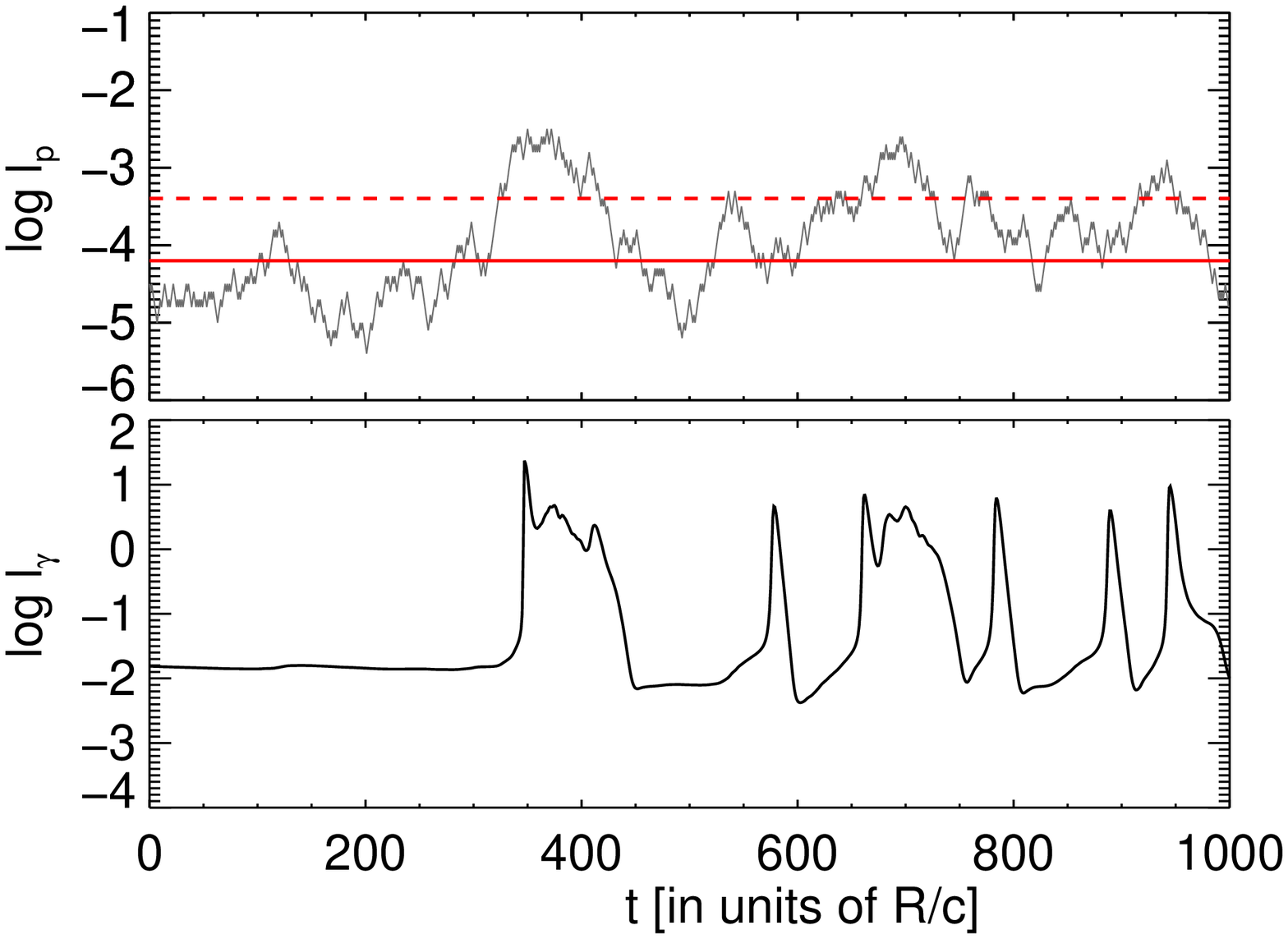}} &
\subfloat[]{\includegraphics[width=0.44\textwidth]{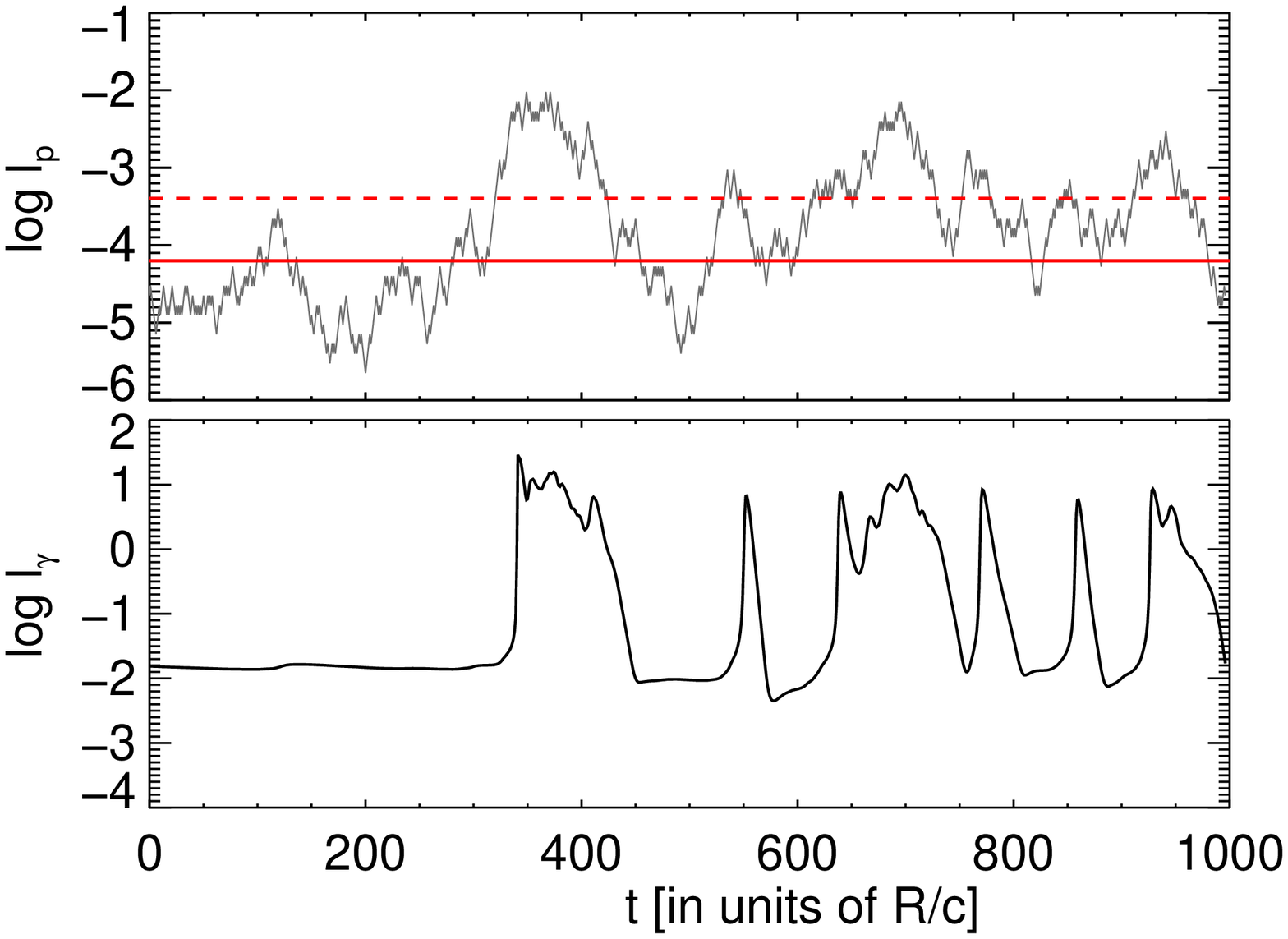}}
\end{tabular}
\caption{Logarithmic plot of the photon compactness as a function of time (in $R/c$ units) obtained for increasing amplitude $A$ -- see eq.~(\ref{eq:red}). Panels (a) to (d) correspond to $A=0.05$, 0.075 , 0.1 and 0.125. The time profile of $\lp$  multiplied by the ratio $m_{\rm p}/m_{\rm e}$ is shown in every case with grey lines. Other parameters same as in Fig.~\ref{white}.}
\label{red}
\end{figure*}

Motivated by the large differences between the temporal profiles of the proton (input) and photon (output) compactnesses, we computed the respective power spectral densities (PSD) by performing a fast Fourier transform on the mean-subtracted normalized (with respect to the maximum value) $\lp(\tau)$ and  $\lph(\tau)$. Figure \ref{psd} shows the logarithmically binned PSDs for three indicative cases shown in Fig.~\ref{white}(a), Fig.~\ref{red}(b), and Fig.~\ref{red}(d). A steepening of the photon PSD at high frequencies is evident, especially in the leftmost case where all the ``flickering'' occurring on short timescales in the proton injection has disappeared from the photon light curve due to the long synchrotron cooling timescale of protons. On the other hand, in the rightmost figure
the photon PSD retain the basic shape of the corresponding proton PSD because, as can be seen from Fig.~\ref{red} the photon light curve mimics the proton one especially during maxima.

\begin{figure*}
 \centering 
 \includegraphics[width=0.32\textwidth]{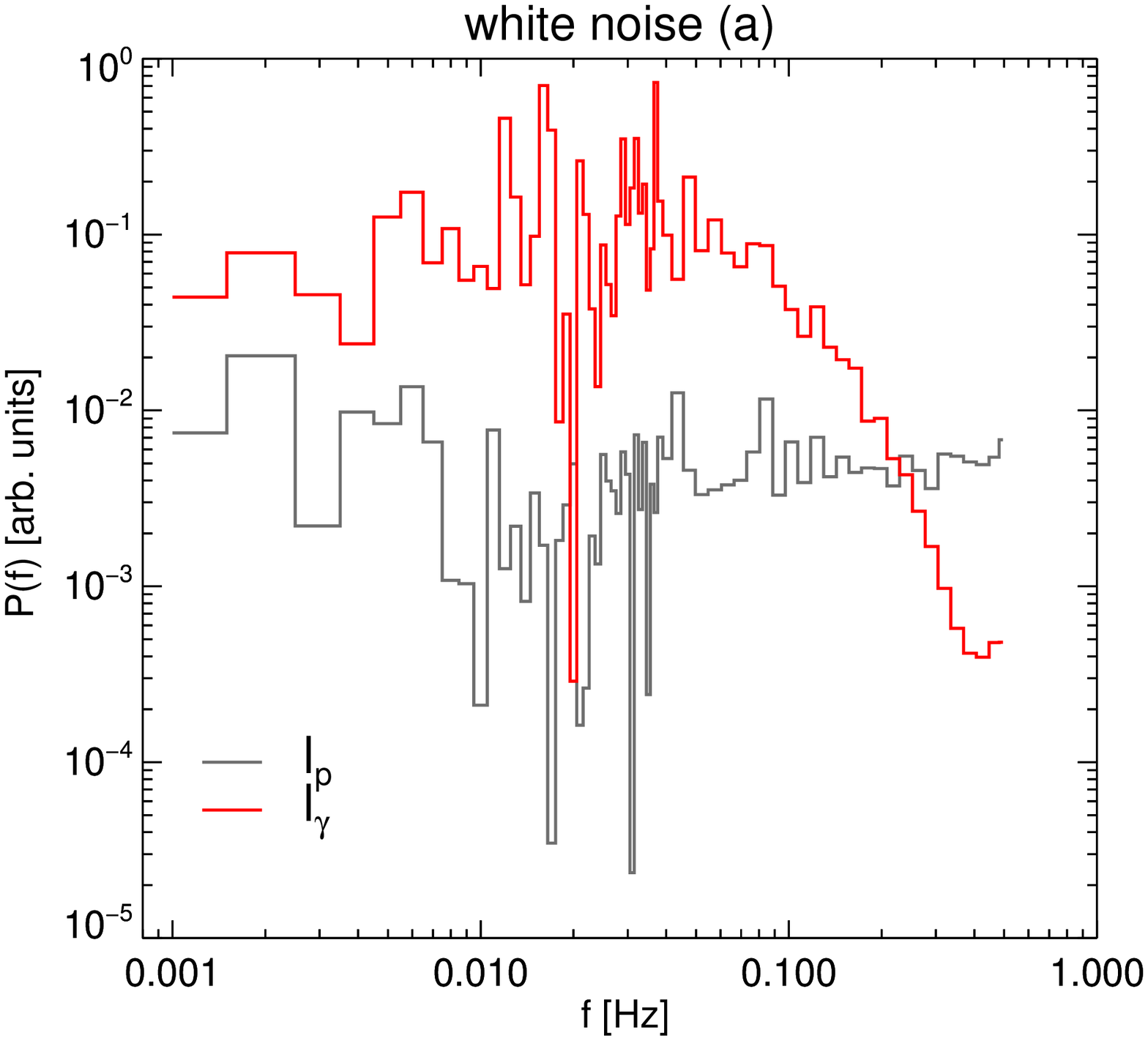} 
 \includegraphics[width=0.32\textwidth]{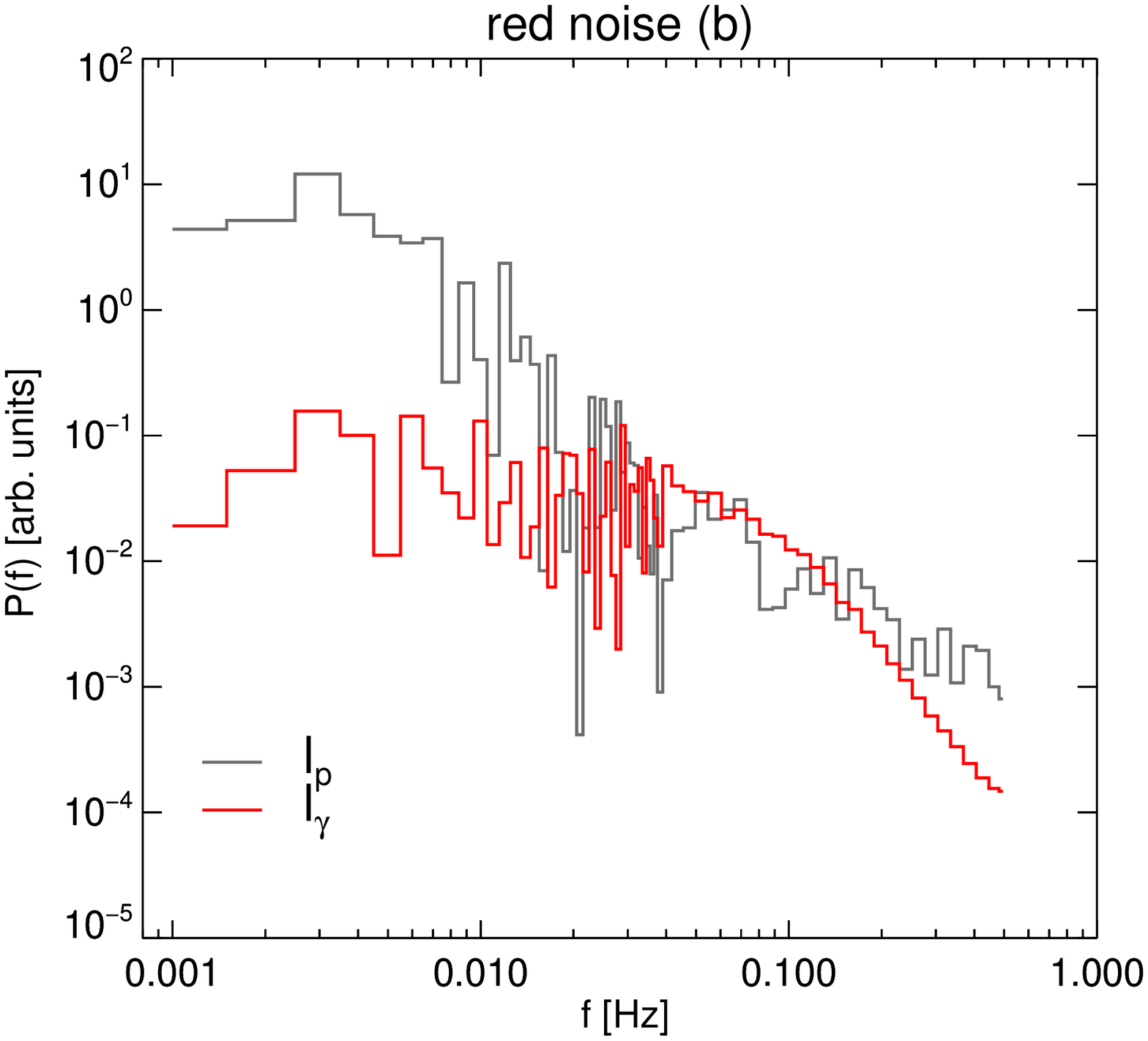} 
 \includegraphics[width=0.32\textwidth]{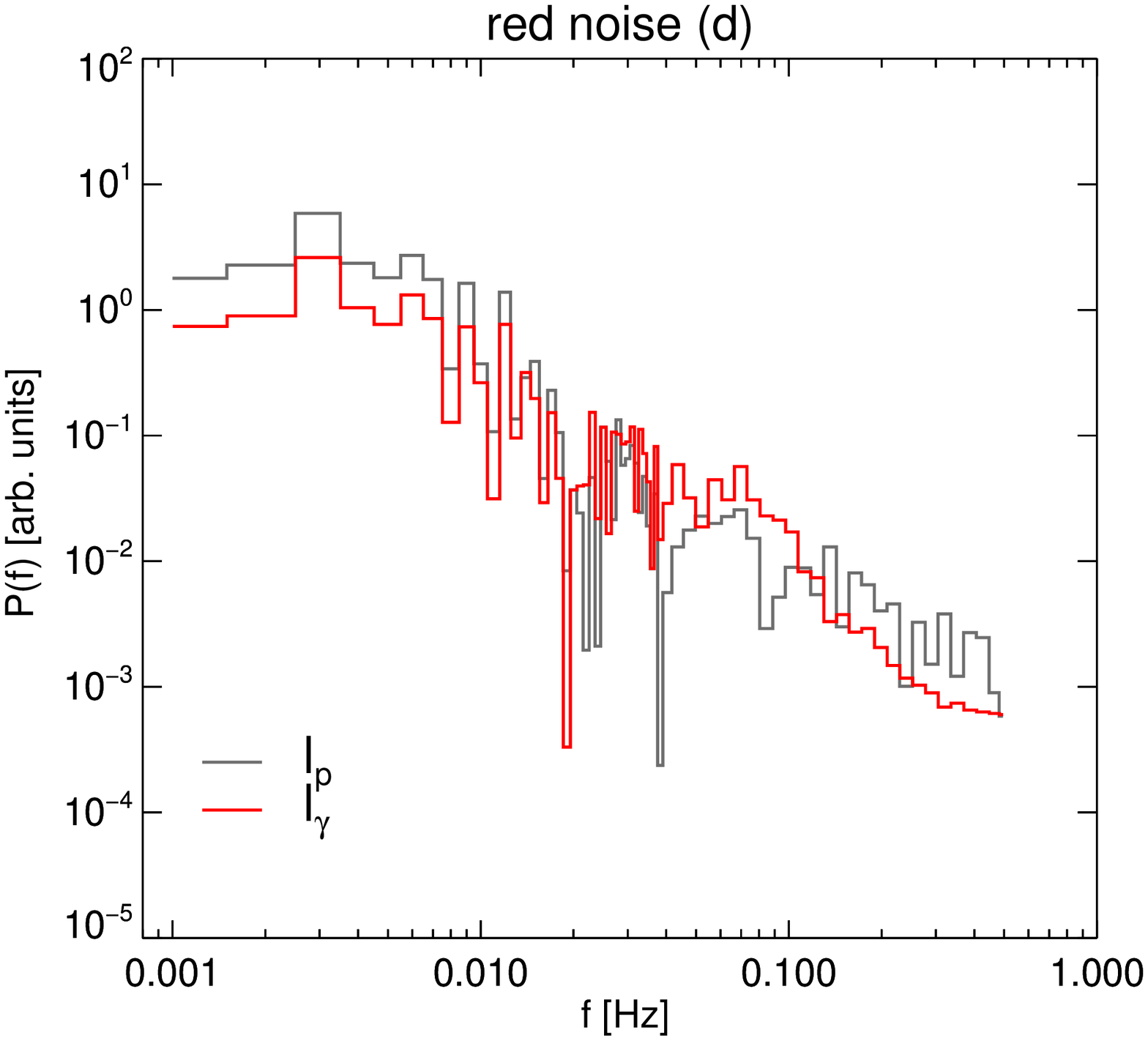} 
 \caption{Logarithmically binned power spectral densities of $\lp$ (grey lines) and $\ell_{\gamma}$ (red lines) for the same cases shown in Fig.~\ref{white}(a),  Fig.~\ref{red}(b), and Fig.~\ref{red}(d) (from left to right).} 
 \label{psd}
\end{figure*}

\section{Relevance to GRB variability}\label{sec:grb}
In this section, we apply our previous results to the GRB 
temporal phenomenology. There is a huge variety of GRB light curves, each of them exhibiting a unique behaviour. Broadly speaking,
they can be divided into two categories: those showing fast-rising-exponential decay (FRED-type)
and those consisting of several sub-pulses with typical widths spanning from hundreds of ms up
to several seconds (e.g. \cite{norris96}). The PSD of observed GRB light curves can be described as a power law, i.e., $P_{\rm f}\propto f^{-\beta}$ with $\beta \approx 5/3$ \citep{belo00,dichiara13}, while there is no compelling evidence for periodicities \citep{guidorzi16}. 
 
We assume a relativistic magnetized outflow with bulk Lorentz factor $\Gamma$. Energy is dissipated internally to the flow at a certain distance from the central engine where particle acceleration to high energies also takes place. The accelerated protons, having a power-law energy distribution of index $s=2$, are injected with luminosity $L_{\rm p}$ into a non-expanding region of size $R$.
The proton compactness is now defined as $\ell_{\rm p}=\sth L_{\rm p}/4\pi R \Gamma^4 \mpr c^3$.  \cite{pdmg14} derived the conditions leading to a proton supercriticality (i.e., the value of $\lpcr$), for a constant energy injection rate. The authors also demonstrated that the photon spectra produced via photohadronic interactions may obtain a Band-like shape due to Comptonization by cooled pairs \citep[for the role of Comptonization in hadronic scenarios, see also][]{murase12}. 

If the GRB central engine shows erratic activity, it is reasonable to assume that the dissipation process is itself variable and, in turn, the proton injection rate. Clearly, as the limit cycle period is a function of the injection, a variable injection will bring the system at various levels of supercriticality  and the superposition of the induced variability with the internal modes of the system can produce some interesting temporal signatures. 

We set to explore the variability patterns using an algorithm  similar to the one introduced in the previous section \citep[see also][]{mpd13}. We assume that the proton injection varies in a random-walk way,  i.e.,  the injection rate is allowed to increase or decrease from one time step to the next one by a small percentage -- see also eq.~(\ref{eq:red})). The parameter values (e.g., magnetic field strength, size, and maximum proton energy) used here are the same as in  Sect.~\ref{sec:temporal}. 
We computed the evolution of the photon flux with time for six proton injection profiles that differ in the  amount of the total energy injected into the system. Our results are presented in Fig.~\ref{lc_obs} where each panel shows the proton injection profile on logarithmic scale (top) and the  observed bolometric light curve on linear scale (bottom). In all cases, the value of the bulk Lorentz factor used to make the transformation to the observer's frame is  marked on the plot, while the red horizontal lines denote the characteristic values of $\lpcr$ and $\lpcrss$, as determined in Sect.~\ref{sec:const}.
\begin{figure*}
\begin{tabular}{ccc}
\subfloat[]{\includegraphics[width=0.3\textwidth]{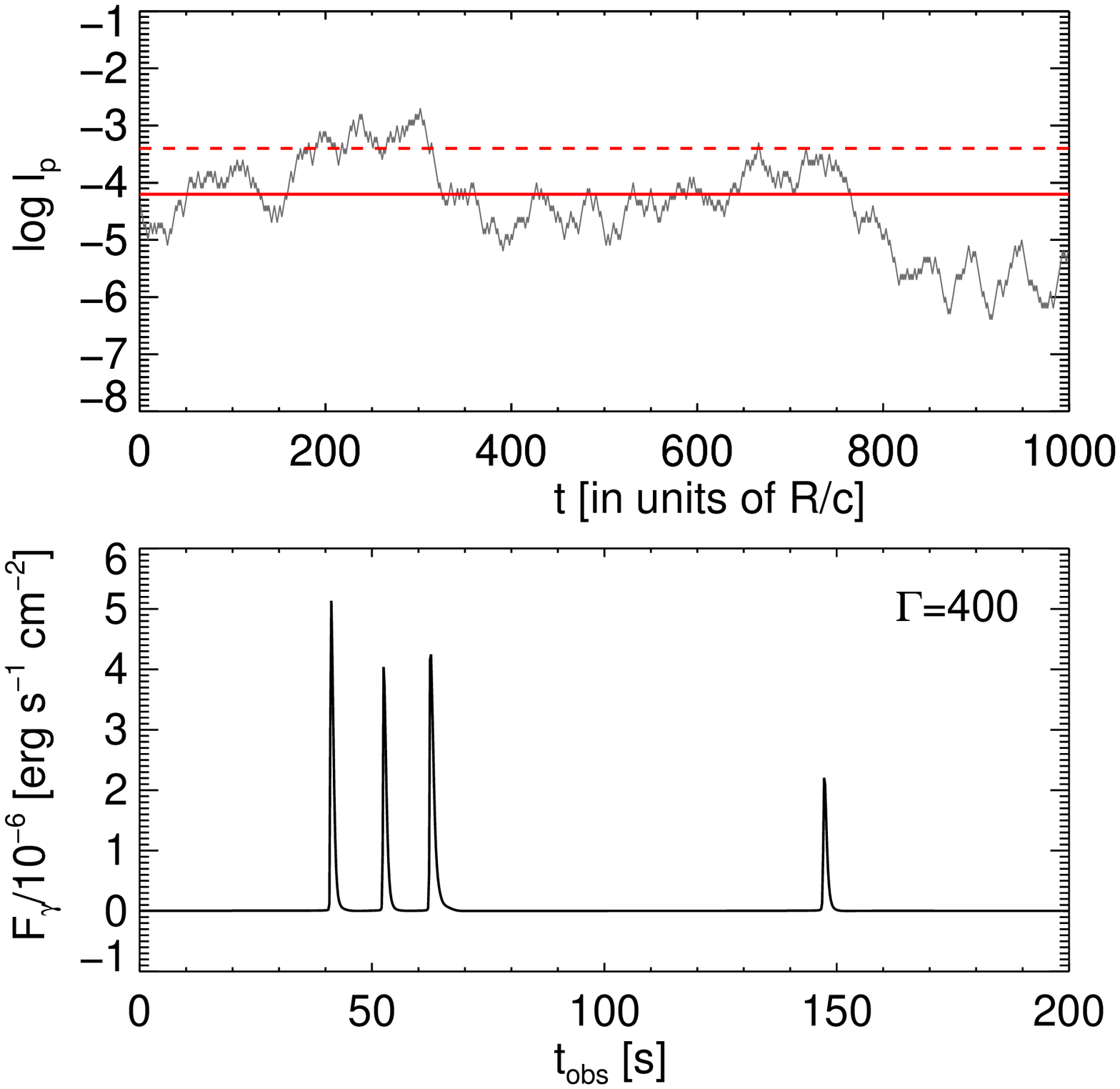}} &
\subfloat[]{\includegraphics[width=0.3\textwidth]{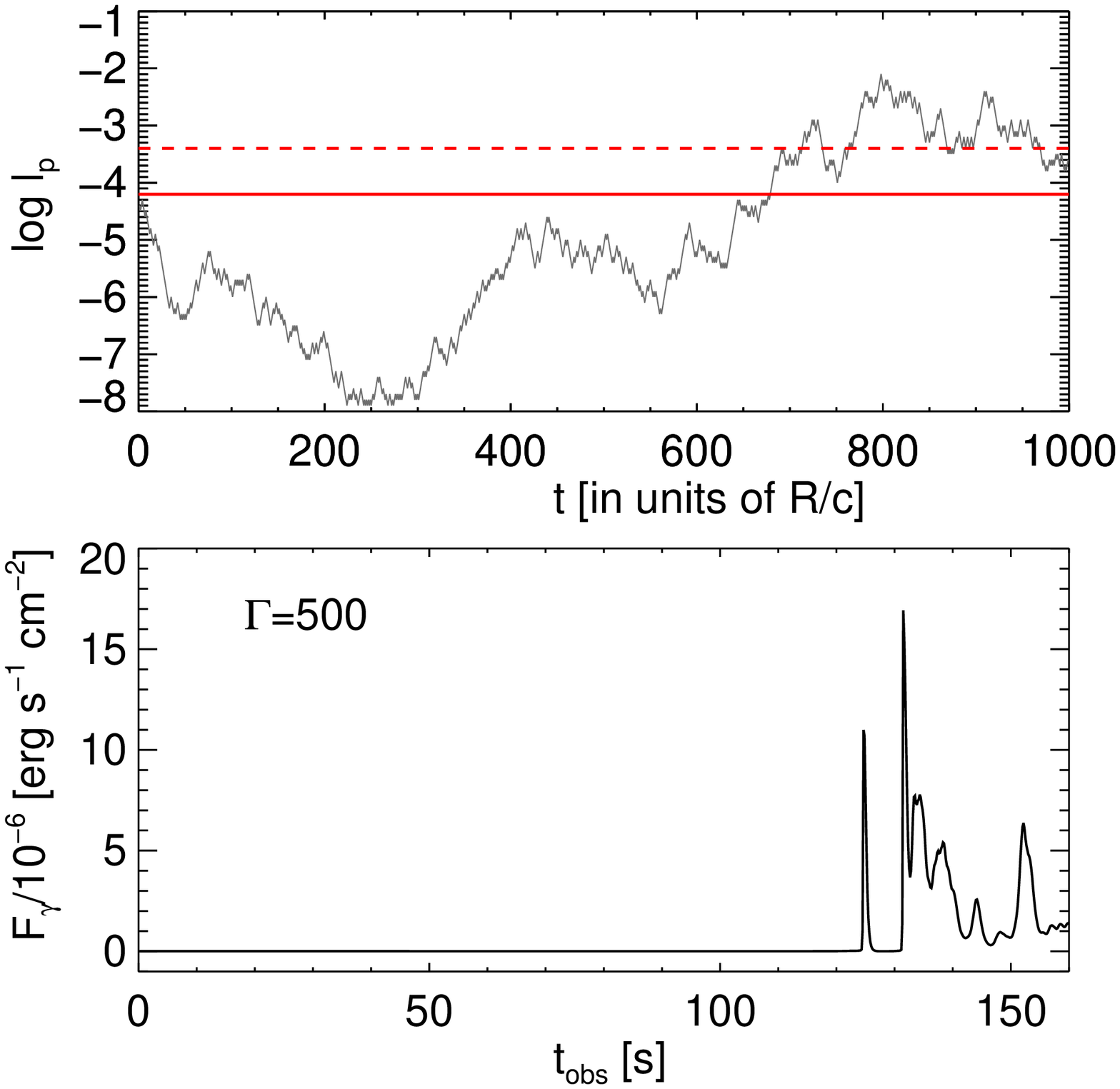}} &
\subfloat[]{\includegraphics[width=0.3\textwidth]{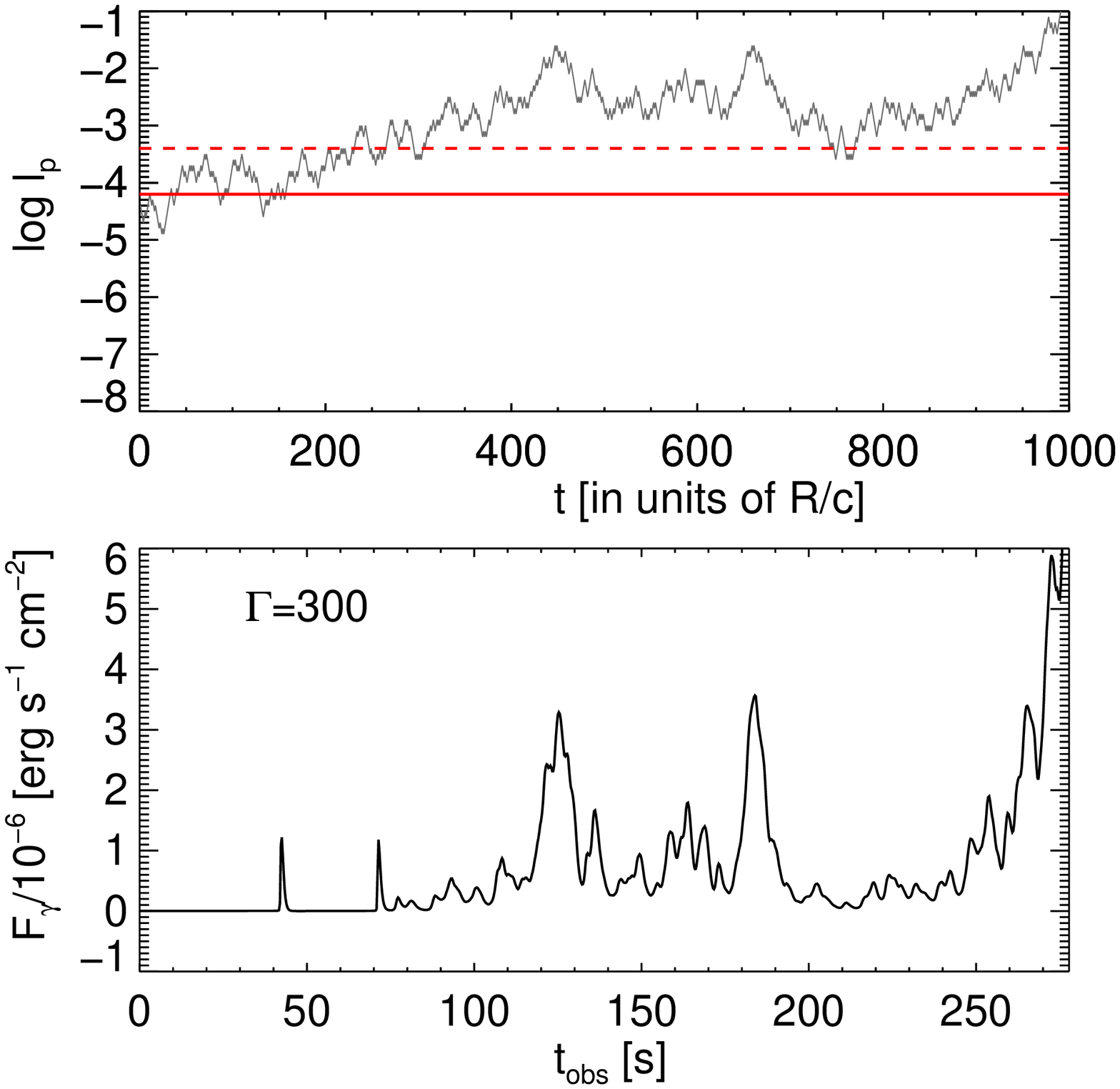}} \\
\subfloat[]{\includegraphics[width=0.3\textwidth]{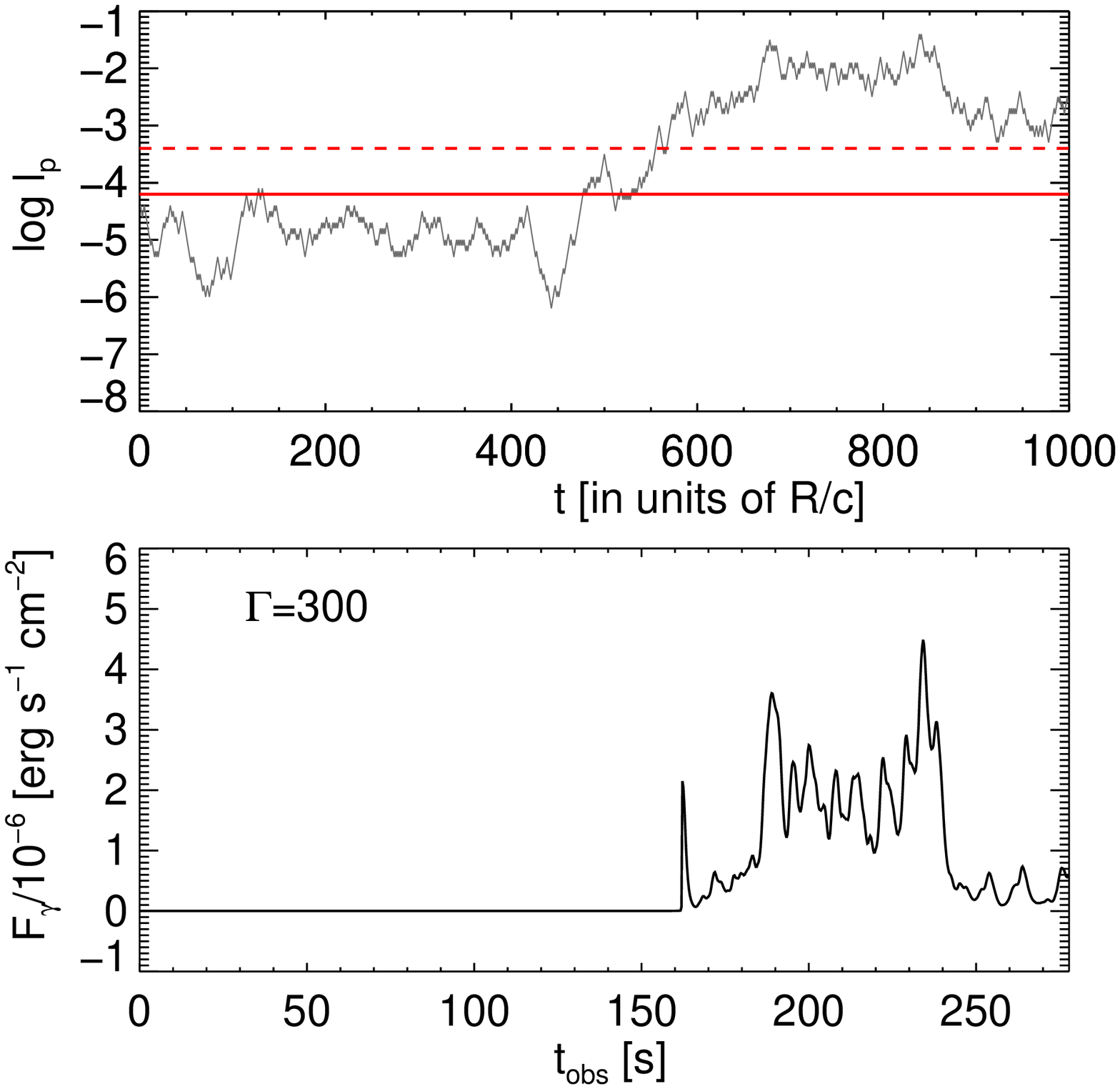}} &
\subfloat[]{\includegraphics[width=0.3\textwidth]{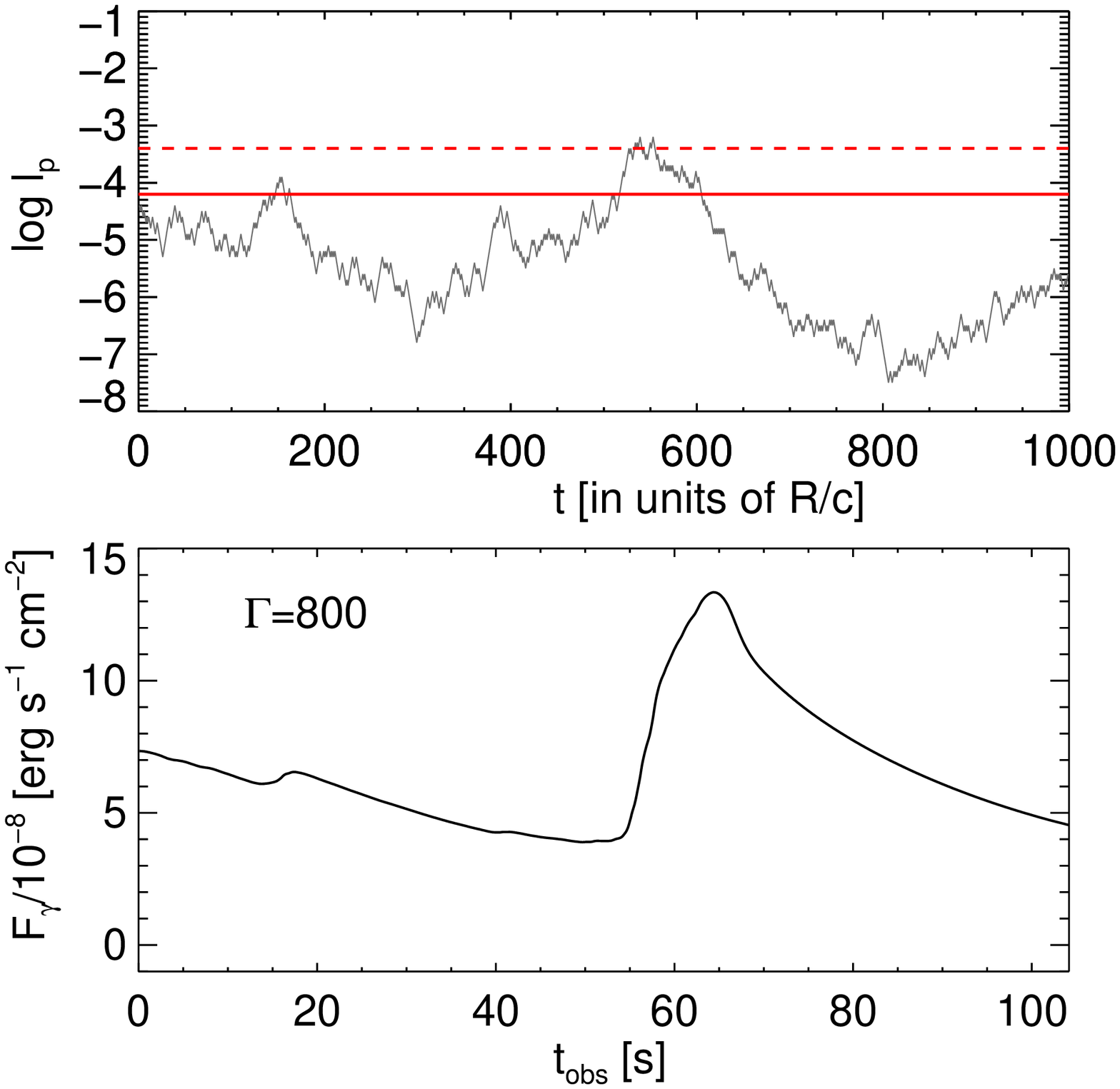}} &
\subfloat[]{\includegraphics[width=0.3\textwidth]{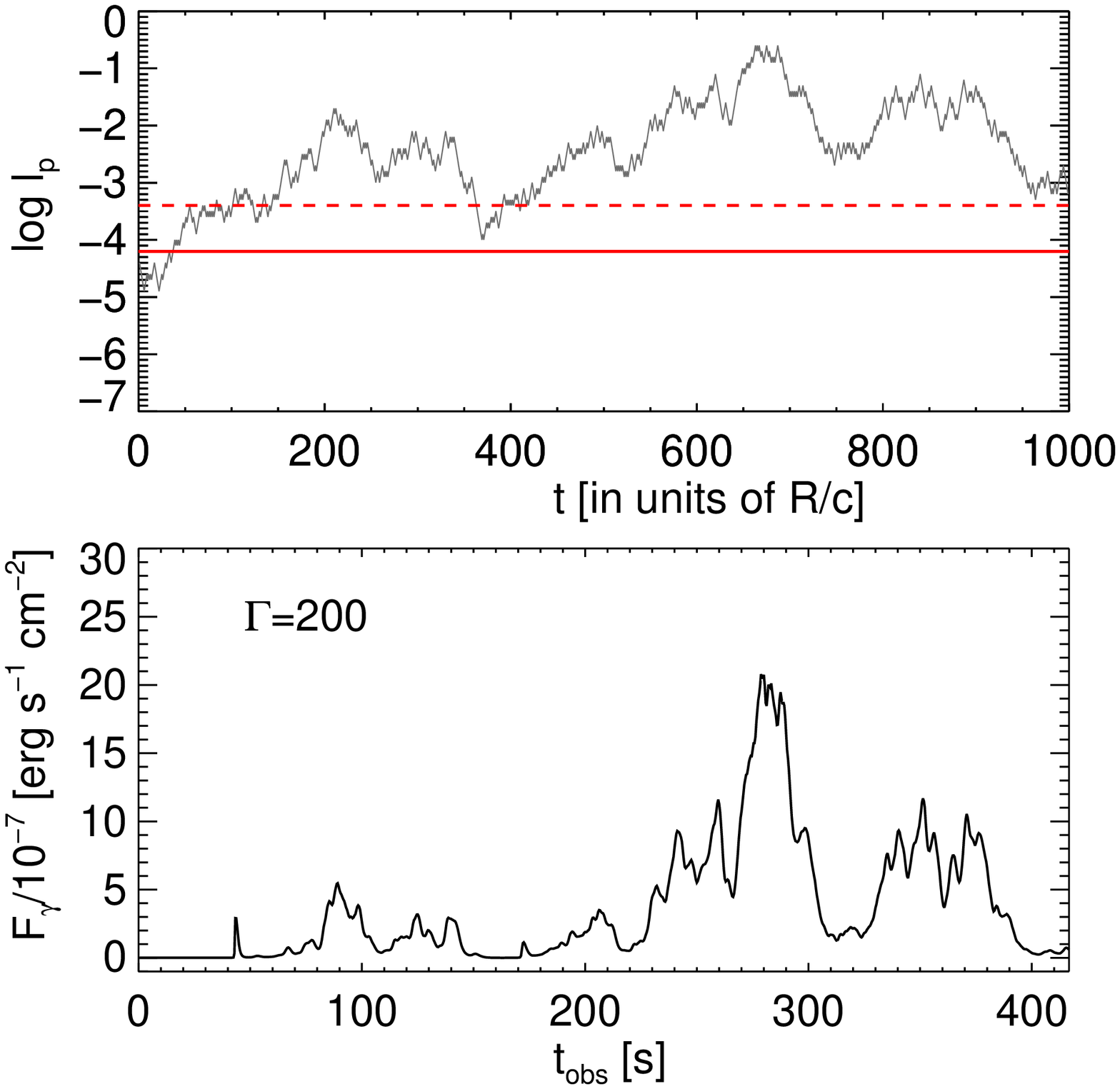}}
\end{tabular}
\caption{Each panel shows the bolometric photon light curve as measured in the observer's frame (bottom) obtained for a proton injection profile (top) whose PSD is $P(f) \propto f^{-\alpha}$, with $\alpha=1.8$.  The values of $\Gamma$ used for the transformation to the observer's frame
are marked on each plot. The source is assumed to be located at a redshift $z=1.5$.}
\label{lc_obs}
\end{figure*}  

Even within the small sample of study cases, we find a wide variety in the obtained light curves, which exhibit (i) single spikes separated by quiescent time intervals (panels a to c),  longer duration pulses with sub-structure (panels c, d, and f), and (iii) a single smooth exponentially decaying pulse (panel e). An one-to-one comparison of $\lp(\tau)$ and $F_{\gamma}(t_{\rm obs})$ shows that no variability is observable whenever $\lp < \lpcr$. This is expected, since in the subcritical regime the radiative efficiency of the system is very low, while the photon spectrum is very different than the Band-like spectrum of typical GRBs \citep[for details, see][]{pdmg14}. However, as soon as $\lp > \lpcr$ (supercritical regime) the radiative efficiency becomes high and the light curve exhibits variability on a range of timescales. For the adopted values of $R$ and $\Gamma$, the variability timescale in the observer's frame -- defined as $\delta t = (1+z) R/c\Gamma$ -- ranges between 0.1~s and 0.5 s. Even shorter timescales may obtained for smaller sources and/or larger Lorentz factors. We also note that in the computed light curves there is no evidence for a correlation between the duration of the pulses and their time of occurrence. This is a result of our assumption that the size of the source remains constant during the injection episode. The bolometric fluences range between $7\times10^{-6}$~erg cm$^{-2}$ (panel a) and $7\times10^{-5}$~erg cm$^{-2}$ (panel c). These values are,  in good approximation, equal to the gamma-ray fluences, since the photon spectrum peaks in the MeV regime in observer's reference frame \citep{pdmg14}. 
 
{In order to facilitate the comparison with the results of our theoretical investigation (see Sect.~\ref{sec:model}), we have chosen the duration of the whole injection episode to be the same (i.e., $T=10^3 R (1+z)/(c\Gamma)$ in the observer's frame).}
The proton energy injected inside the blob  during this period $T$ can be calculated as $4 \pi R m_p c^3 \Gamma^4 \sigma_{\rm T}^{-1} \int_{0}^{T} dt_{\rm obs} \ell_{\rm p}(t_{\rm obs})$. The bolometric photon energy released during the same time interval $T$, $E_{\gamma}$, is a significant fraction of the proton energy, as shown in Table~\ref{tab:eff}. The radiative efficiency $\eta \equiv E_{\gamma}/E_{\rm p}$ is higher than 20 per cent in all cases, but one (run e). In the latter case, the system becomes supercritical only for a short time window during the whole episode, thus leading to a reduced efficiency (see Fig.~\ref{lc_obs}.)

\begin{table}
\centering
\caption{Total (isotropic) injected proton energy $E_{\rm p}$ and bolometric photon energy $E_{\gamma}$ (in the observer's frame) computed for the six cases presented in Fig.~\ref{lc_obs}. The last column shows the radiative efficiency $\eta$.}
\label{tab:eff}
\begin{tabular} {ccccc}
Run  & $\Gamma$ & $E_{\rm p}$ [erg] & $E_\gamma$ [erg] & $\eta$ \\
\hline
 a  & 400 & $6.8\times 10^{53}$ & $2.3 \times 10^{53}$ & 0.33 \\
 b & 500 & $3.6 \times 10^{54}$ & $1.4 \times 10^{54}$ & 0.38 \\
 c & 300 & $10^{55}$ & $3\times10^{54}$ & 0.30\\
 d & 300 & $5.5\times 10^{54}$ & $2.1\times 10^{54}$ & 0.38 \\
 e & 800 & $10^{54}$ & $9.9\times 10^{52}$ & 0.09 \\
 f & 200 & $9.3\times10^{54}$ & $2.1\times10^{54}$ & 0.23 \\
 \hline
\end{tabular}
\end{table}

The PSDs of the light curves\footnote{We performed the fast Fourier transform on the mean-subtracted normalized (with respect to the maximum value) light curves.} shown in Fig.~\ref{lc_obs} are presented in Fig.~\ref{psd_obs}. The PSDs were binned with a logarithmic bin of 1.1. They can be described by a single or a broken power law (panels b-f and panel a, respectively). We find that the power-law index of the PSD ranges between $\sim -2.5$ to $\sim -1$, while 
the PSD of the proton injection profile is, in all cases, $P(f) \propto f^{-\alpha}$ with $\alpha=1.8$. Red lines with slopes -1 (solid), -2 (dashed), and -3 (dotted) are overplotted in all panels for comparison. The PSD of the light curve does not always resemble that of the injection profile, as illustrated in panels (a) and (f). In the former case, for example, the radiation is released in a series of bursts that are separated in time, although the energy injection is a red noise process. 

\begin{figure*}
\begin{tabular}{ccc}
\subfloat[]{\includegraphics[width=0.3\textwidth]{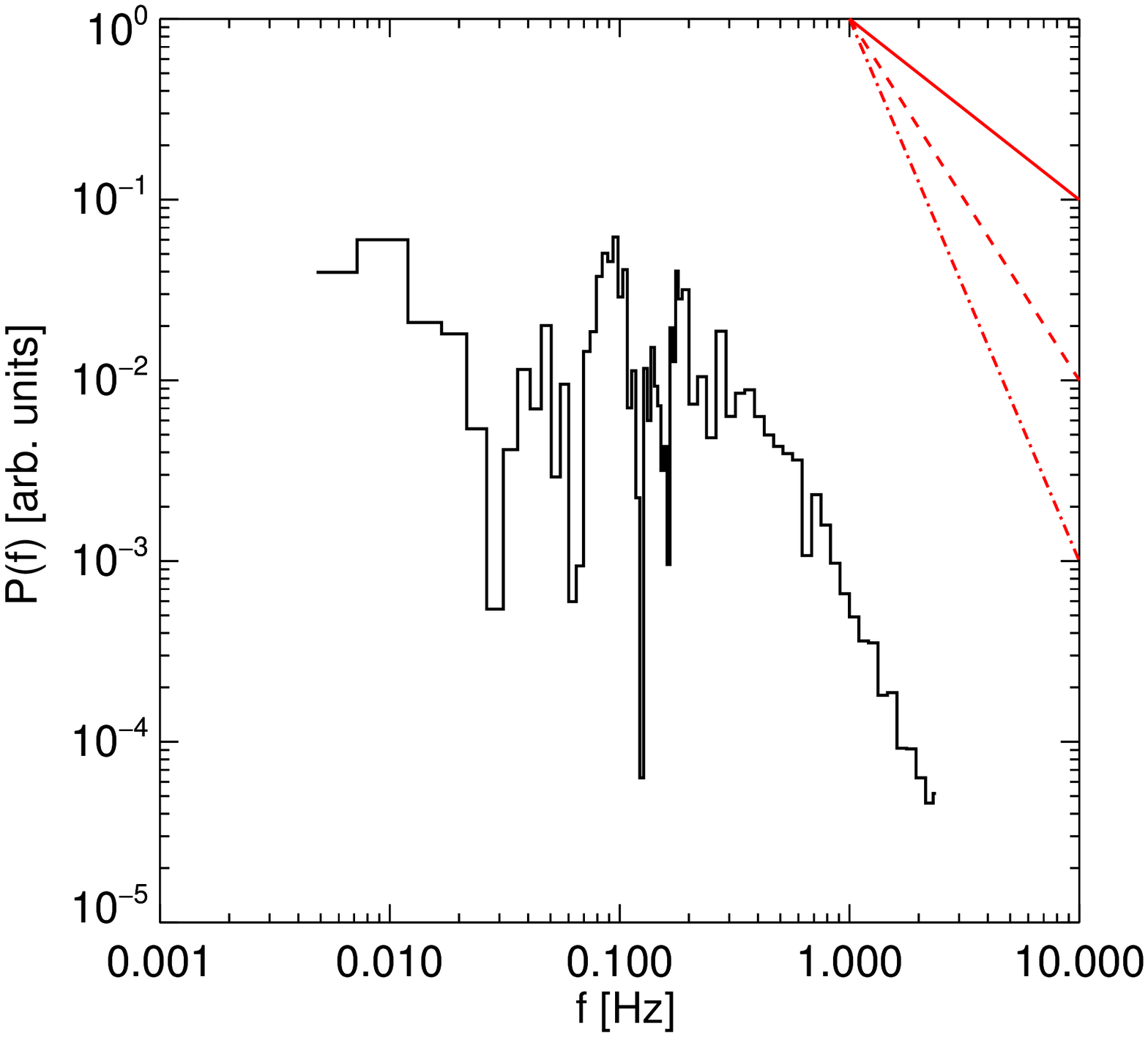}} &
\subfloat[]{\includegraphics[width=0.3\textwidth]{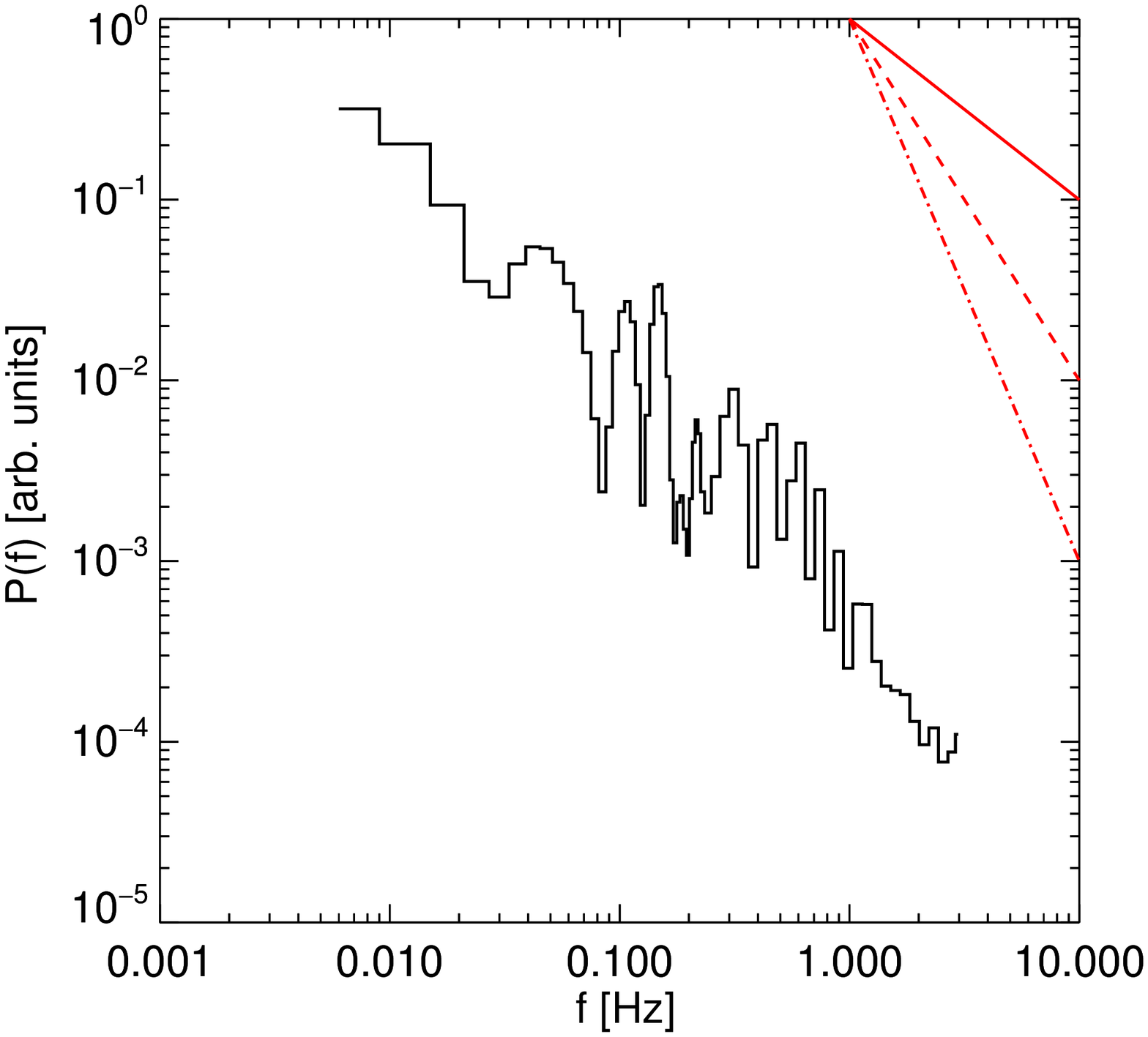}} &
\subfloat[]{\includegraphics[width=0.3\textwidth]{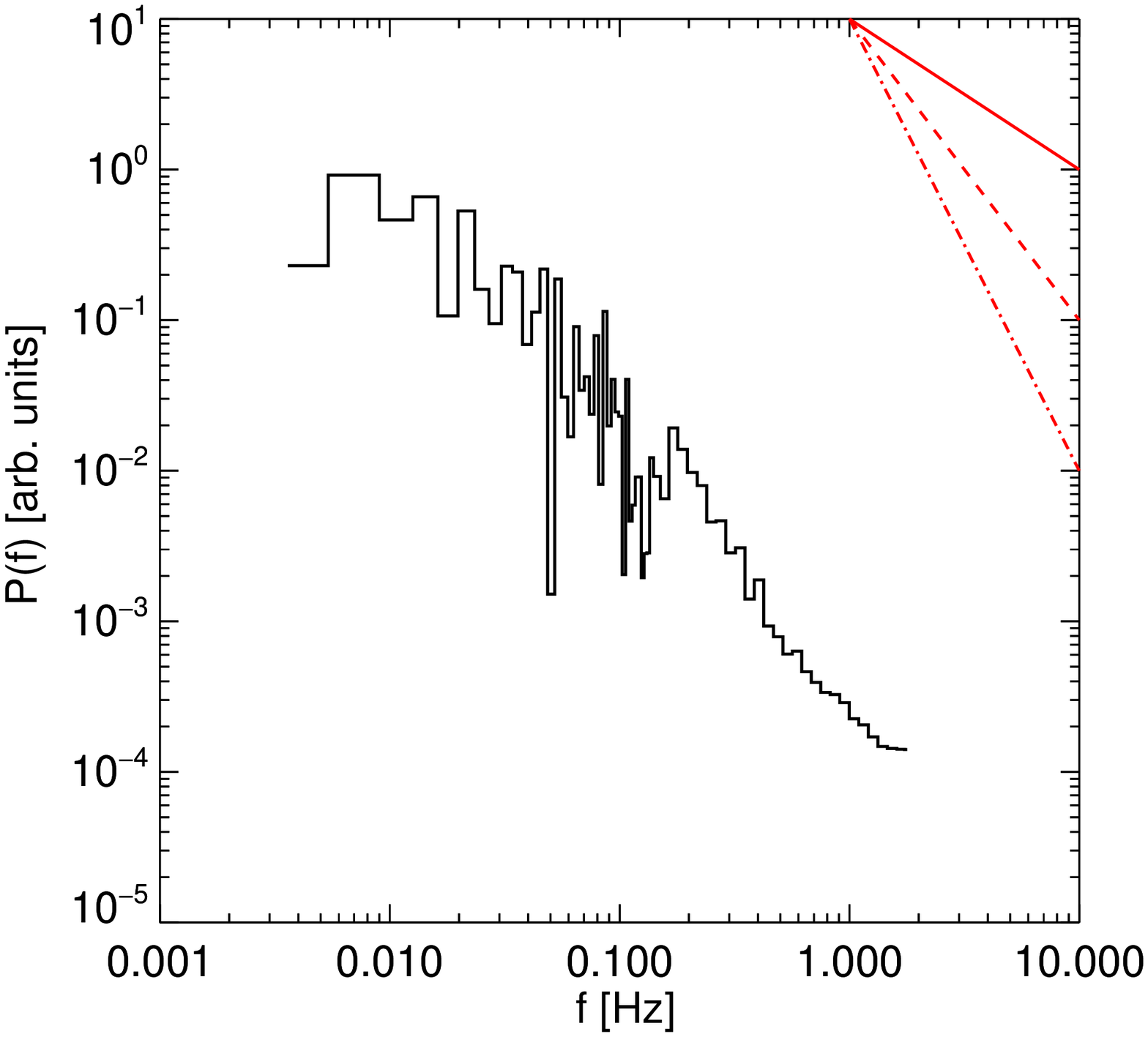}} \\
\subfloat[]{\includegraphics[width=0.3\textwidth]{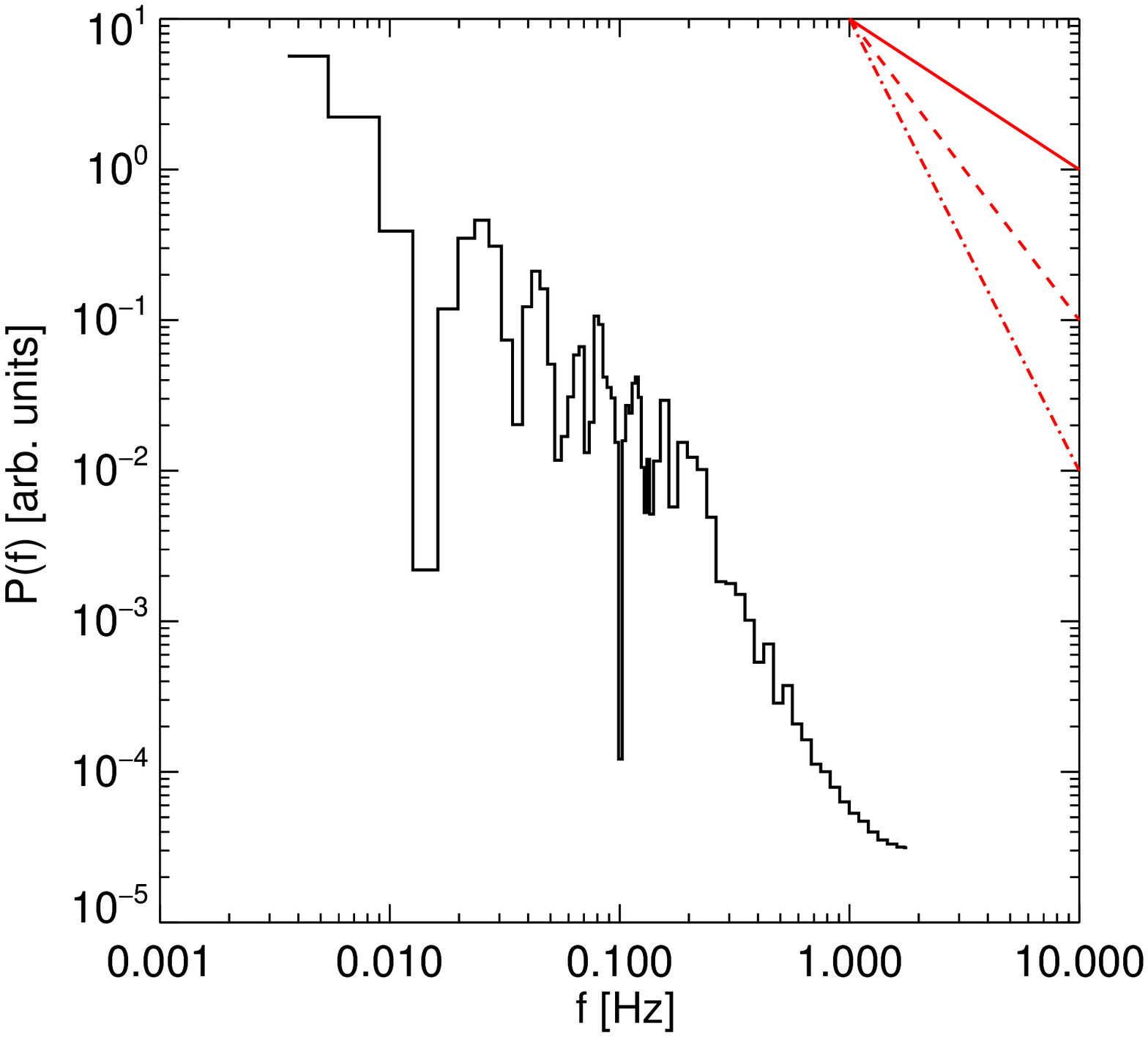}} &
\subfloat[]{\includegraphics[width=0.3\textwidth]{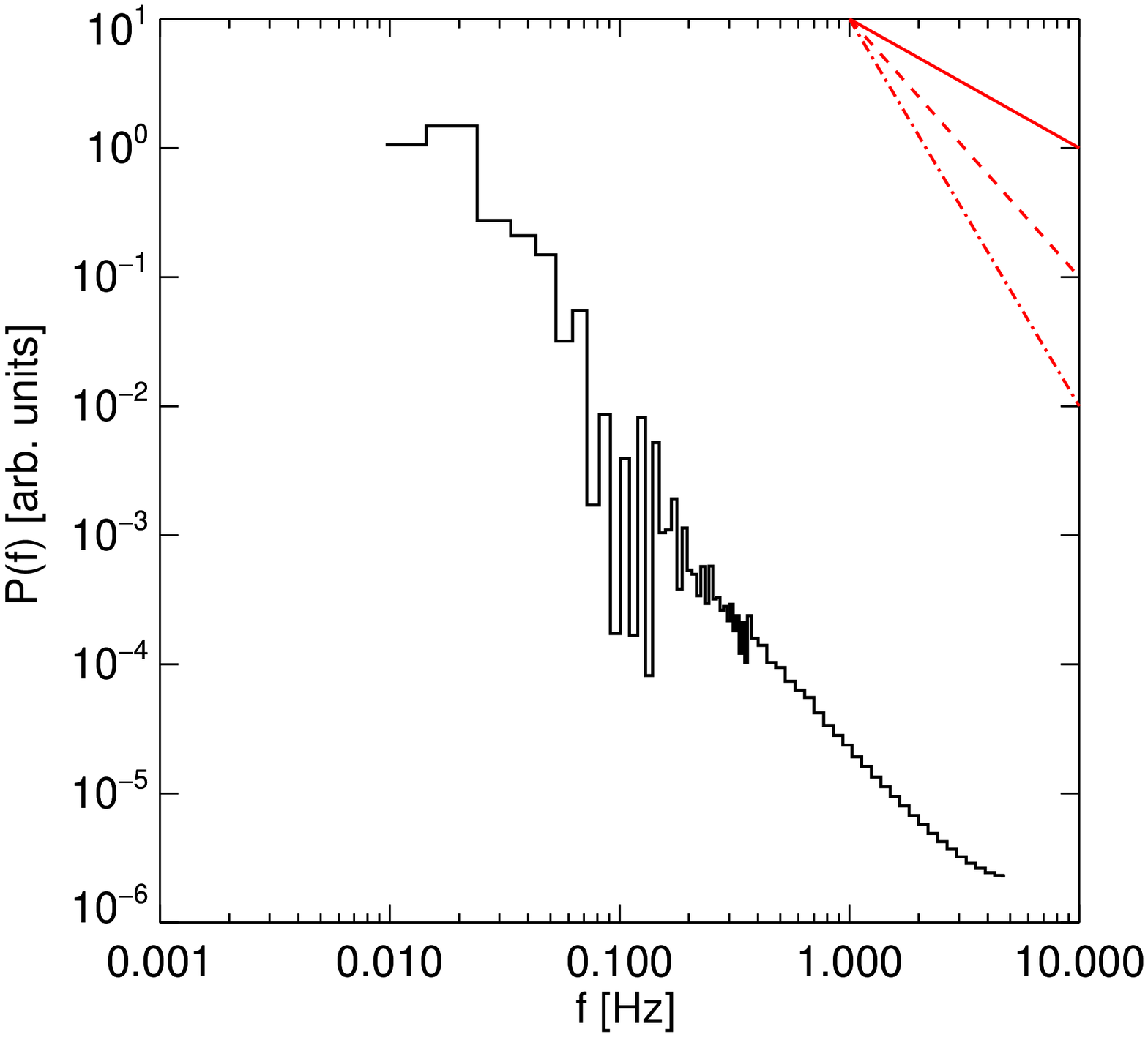}} &
\subfloat[]{\includegraphics[width=0.3\textwidth]{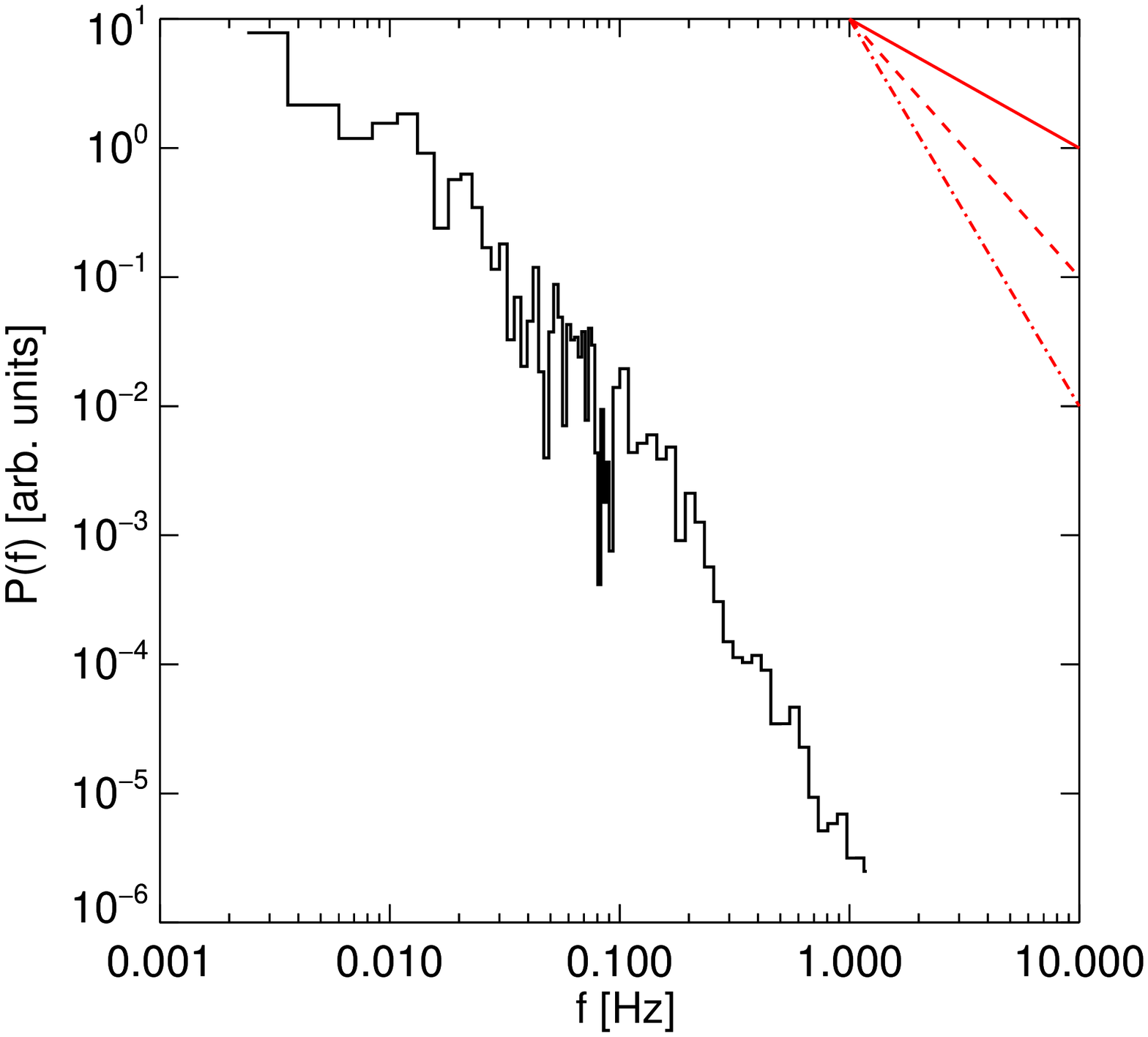}}
\end{tabular}
\caption{Logarithmically binned PSDs of the observed bolometric photon light curves shown in Fig.~\ref{lc_obs}. Red lines with slopes -1 (solid), -2 (dashed), -3 (dotted) are overplotted for comparison. 
In all cases, the PSD of the proton injection profile is $P(f) \propto f^{-\alpha}$ with $\alpha=1.8$. }
\label{psd_obs}
\end{figure*}

\section{Discussion}\label{sec:discuss}

In the present paper we have examined the variability patterns expected in the context of one-zone hadronic models. As it was shown by \cite{kirkmast92} when the density
of the relativistic protons is above some critical value, then the system
becomes supercritical and exhibits some interesting temporal behaviour.

We first investigated the dynamical behaviour of the system in the case of  constant energy injection rate and verified the results of previous studies \citep{sternsvensson91, mastetal05, petromast12b}. These are summarized below:
\begin{itemize}
 \item for low enough proton injection compactnesses, a subcritical steady state is established. In the absence of external photons, the main loss mechanism for protons is  synchrotron radiation and the system can be very inefficient especially when $\gmx$   and $B$ are such that  the synchrotron cooling timescale is much longer than the dynamical one. 
 \item if the proton compactness exceeds a critical value $\lpcr$, then the system becomes supercritical, i.e. the energy stored into relativistic protons is released within a few dynamical timescales. For the lowest values of   $\lp$ that are still above $\lpcr$, the system exhibits quasi-periodic oscillations (limit cycles) that become more frequent as $\lp$ increases. The reason is that the ``burned'' proton energy  in each photon outburst is replenished at a higher rate as $\lp$ increases. This, in turn, causes the outbursts to occur in shorter time intervals. 
 \item for even larger values of $\lp$ above $\lpcr$, the system shows a dumped oscillatory behaviour. In this regime, the  proton energy 
is replenished so fast that the light curve reaches a steady state. This supercritical steady state is characterized by strong photopair and photopion  losses for the  protons with synchrotron contributing only a small percentage. 
This behaviour is common almost for all initial parameters, i.e. in most cases the transition of a system
 to supercriticality  is achieved through a succession of limit cycle states.
  If we denote $\lpcrss$ the value of the injected compactness where the system reaches quickly -- i.e. after one or two outbursts  -- the supercritical steady state,  we typically find that $\lpcrss \sim 10\ \lpcr$.
\end{itemize}

As a second step, we investigated the variability patterns of the hadronic system in the general case of a variable energy injection rate. The basic findings of our analysis (see Sect.~\ref{sec:temporal}) can be summarized in the following points:
\begin{enumerate} 
 \item the photon light curves tend to follow closely the variations of the injection 
 either when the system is subcritical ($\lp < \lpcr$ or low zone) or when the conditions drive it
 deep in the supercritical regime ($\lp > \lpcrss$ or high zone). 
 \item the  intermediate zone, which is a unique feature of
the leptohadronic systems, acts in the exact opposite way 
of a typical frequency filter by allowing the emergence of specific frequencies.
 \item the photon light curves exhibit complex patterns whenever the injection 
 falls in the intermediate zone ($\lpcr \le \lp \le \lpcrss$). The observed structured flares
 are the result of the superposition of the intrinsic periodicity and the variability of the energy injection source. 
\item there is not a one-to-one relation between the power spectral density of the energy injection function 
and that of the photon light curves. 
\item the slope of the power spectral density of photon light curves is not sensitive on the 
total injected energy but rather on the time spent in the intermediate and high zones. 
\end{enumerate}

The work presented here can be considered as a continuation of \cite{pdmg14} who showed that for parameters relevant to GRBs, the system produces photon luminosities and spectra that are very similar to those observed during the GRB prompt emission phase. In this paper, we show that another major property of hadronic systems, i.e. their inherent non-linearity, can be applied to GRBs that are clearly dynamical systems. However, to fully apply this model to the emission of GRBs, one has to provide a physical mechanism that is responsible for the variable particle injection \citep[for other scenarios for GRB variability, see e.g.][]{Z14}.}

In the present treatment we have deliberately ignored the presence of accelerated electrons in the source, since we wanted to study the temporal properties of the system with as few free parameters as possible. The presence of electrons in the source would result in the production of additional photons (e.g. synchrotron) that  could, in principle, act as targets for photohadronic interactions. Thus, relativistic electrons could affect the onset of supercriticality by lowering the critical proton compactness $\lpcr$. For example, the injection of accelerated electrons with a power law distribution of slope $s=2$ extending to $\gamma_{e,\max}=\gamma_{p,\max}\left(m_{\rm e}/m_{\rm p}\right)$ with luminosity $L_{\rm e}=0.01 \ L_{\rm p}$ would correspond to a 3\% lower value of $\lpcr$ for the same parameters as those used in Fig.~\ref{Qcon}. We also find that the frequency of quasi-periodic oscillations increases with increasing electron luminosity. When the latter exceeds  $\sim 25$\% of the proton luminosity, the intrinsic variability of the system is washed out and the system reaches a steady state as soon as it becomes supercritical.

The onset of hadronic supercriticality depends on  a combination of physical parameters, such as the maximum proton energy and the magnetic field strength \citep[e.g.][]{kirkmast92, petromast12}. Yet, the variability patterns discussed here are not sensitive to the specifics of the model. For instance, our results do not depend on the duration of the energy injection {\sl per se}, but rather on the time that the system spends on the supercritical regime.  If the emitting region was smaller (e.g., $R=10^{11}$~cm), we would still obtain similar results as those shown in Fig.~\ref{lc_obs} except for the shorter variability timescale and total duration of the episode.

The variability signatures from the source do not rely on the presence of ultra high-energy protons in the source. The neutrino signal predicted for different types of light curves and different maximum proton energies is worth investigating \citep[see also][for neutrino emissivity based on various types of GRB variability] {bustamante17}. For  sufficiently low proton energies that ensure the onset of supercriticality, the produced neutrinos might have enough low energies  (e.g. $<100 $~TeV) that the constraints put on GRBs as neutrino emitters \citep[e.g.][]{aartsen15, aartsen16} would not apply. We plan to address this issue in a future publication.

Summarizing,  systems containing magnetic fields, relativistic protons, and pairs can exhibit unique variability patterns as a result of a non-linear network of the radiative processes in play. This paper presents a new aspect of hadronic variability with potential application to compact high-energy emitting sources.

\section*{Acknowledgements}
We thank the referee for their constructive report. MP acknowledges support by the L. Spitzer Postdoctoral Fellowship. We thank G. Vasilopoulos for useful discussions.

\bibliographystyle{mnras} 
\bibliography{grbhadro.bib} 

\end{document}